\documentclass[fleqn,usenatbib]{mnras}

\usepackage{newtxtext,newtxmath}
\usepackage[T1]{fontenc}

\DeclareRobustCommand{\VAN}[3]{#2}
\let\VANthebibliography\thebibliography
\def\thebibliography{\DeclareRobustCommand{\VAN}[3]{##3}\VANthebibliography}

\usepackage{graphicx}	% Including figure files
\usepackage{amsmath}	% Advanced maths commands
\title[Caustic crossing events]
 {On the detection of caustic crossing events associated with dark
 matter in the form of primordial black holes}

\author[M. R. S. Hawkins]{
M. R. S. Hawkins $^{1}$\thanks{E-mail: mrsh@roe.ac.uk}
\\
$^{1}$Institute for Astronomy (IfA), University of Edinburgh,
 Royal Observatory, Blackford Hill, Edinburgh EH9 3HJ, UK\\}

\date{Accepted XXX. Received YYY; in original form ZZZ}

\pubyear{2022}

\begin{document}
\label{firstpage}
\pagerange{\pageref{firstpage}--\pageref{lastpage}}
\maketitle

\begin{abstract}
The possibility that stellar mass primordial black holes may make up at
least a significant fraction of dark matter has recently received much
attention, partly as a result of gravitational wave observations, but more
specifically from observations of microlensing in the Galactic halo and in
quasar gravitational lens systems.  If this is the case then a number of
observable consequences are to be expected.  This paper focusses on the
prediction that dark matter in the form of primordial black holes will
result in a web of caustics which when traversed by quasars will result in
a complex but characteristic amplification of the accretion disc light
source.  Caustic crossings produce features in quasar light curves which
are relativly straightforward to identify, and are hard to associate with
any intrinsic mode of variation.  Microlensing simulations are used to
clarify the nature of the expected light curve features, and compared with
observed light curves to demonstrate that caustic crossing features can be
present.  A further test of microlensing is based on the expected
statistical symmetry of the light curves, which is not predicted for most
models of intrinsic quasar variability, but is found in large samples of
quasar light curves.  The conclusion of the paper is that observations of
quasar light curves are consistent with the expected microlensing
amplifications from dark matter made up of stellar mass primordial black
holes, but cannot easily be explained by intrinsic variations of the
quasar accretion disc.
\end{abstract}

\begin{keywords}
quasars: general -- gravitational lensing: micro -- dark matter
\end{keywords}

%%%%%%%%%%%%%%%%%%%%%%%%%%%%%%%%%%%%%%%%%%%%%%%%%%

%%%%%%%%%%%%%%%%% BODY OF PAPER %%%%%%%%%%%%%%%%%%

\section{Introduction}
\label{int}

The idea that dark matter may be in the form of primordial black holes,
or perhaps other compact bodies, has received considerable observational
support in the last few years \citep{c24}.  The main obstacle to the
identification of compact bodies as dark matter over the last 50 years or
so has been the widely accepted view that dark matter is in the form of
non-baryonic elementary particles.  The reason for the acceptance of this
paradigm is not entirely clear, as attempts to detect plausible dark
matter particles have consistently failed over this period.  As the
limitation of the neutrino floor or `fog' \citep{o21} is approached by
dark matter particle detectors, the prospects for direct detection do not
look good, and there now seems to be a paradigm shift towards primordial
black holes as dark matter.

If indeed dark matter is in the form of primordial black holes, then the
associated mass function is expected to be very broad, but with some well
defined peaks in probability, notably associated with phase transitions
\citep{c21}.  Of particular interest is the peak associated with the QCD
phase transition \citep{b18b} where the most probable mass for a primordial
black hole is around $0.7 M_\odot$.  If a significant fraction of dark
matter is indeed in the form of solar mass primordial black holes, then
there are a number of cosmological situations where they would be expected
to betray their presence in a detectable way.

The first large scale programme to detect dark matter in the form of
compact bodies \citep{a96,a00} involved monitoring several million stars
in the Magellanic Clouds to look for evidence of microlensing by compact
bodies in the Galactic halo.  The results implied the detection of a large
halo population of stellar mass compact bodies which could not be
accounted for by any known stellar population.  This raised the question
of whether these objects could make up the dark matter.  On the assumption
of a flat halo rotation curve and a relatively narrow mass function, the
authors concluded that less than 40\% of the halo could be made up of
compact bodies \citep{a00}.  However, it is now clear that the Milky Way
rotation curve is not flat \citep{h15,c18}, but declines implying a
reduced dark matter content which is consistent with the compact bodies
detected by the MACHO project making up the dark matter.   Recently, the
decline of the Milky Way rotation speed has been confirmed beyond
reasonable doubt \citep{j23,o24,g24} by observations from the {\it Gaia}
DR3 data release\footnote {https://www.cosmos.esa.int/web/gaia/dr3}.

The detection of a compact body component of dark matter in the Milky Way
halo was followed by investigations into the make-up of more distant
galaxy halos.  In particular, gravitational lens systems have provided a
very useful environment for detecting compact bodies in galaxy halos.  It
is well-known that as well as varying in response to changes in quasar
luminosity, individual images vary independently, often by large amounts
\citep {m09,p12,b18a}.  These changes have been widely attributed to
microlensing by compact bodies in the halo of the lensing galaxy, but
there has been some debate as to whether the lenses are halo stars, or
compact bodies making up the halo dark matter \citep{h20a}.  The question
now seems to be settled by observations of the cluster lens
SDSS J1004+4112 where strong microlensing amplifications are observed some
60 kpc from the cluster centre, far beyond the distribution of stars which
declines to a negligible level beyond 25 kpc \citep{h20b}.  If a
significant fraction of dark matter is indeed in the form of compact
bodies, then there are a number of other situations where one might expect
them to betray their presence.

An early observation that the expected time dilation in quasar light
curves due to intrinsic variations was not observed \citep{h10} led to the
proposal that apparent changes in quasar luminosity were more plausibly
attributed to microlensing amplifications, where the detection of time
dilation would not be expected.  Recently however it has been shown
\citep{l23} that time dilation is seen in quasars, implying
that any time dilation signal from intrinsic variations can still be
detected.  This is consistent with recent work \citep{h22} showing that
to reproduce the observed distribution of quasar amplitudes of variation
it is necessary to include both intrinsic variations and microlensing
amplifications from a cosmic distribution of stellar mass compact bodies.

In addition to amplifying the light from the quasar nucleus, there is a
strong case that knots in a turbulent broad line region will also be
microlensed, leading to rapid changes in the shape and structure of quasar
broad emission lines \citep{h24}.  This has been observed in a number of
quasar spectra, where new components of broad emission lines appear or
disappear on a timescale of a few years, an order of magnitude shorter
than the dynamic timescale of the broad line region \citep{p93}.  This is
best explained by rapid microlensing amplification of knots in the broad
line region with radial velocities differing by several hundred km
sec$^{-1}$ from the systemic velocity, and is consistent with expectations
for a population of primordial black holes making up the dark matter.

There have been many attempts over the years to find evidence for
microlensing, and hence a population of compact bodies, by examining
the properties of quasar light curves.  The main difficulty with this
approach has always been to distinguish microlensing from intrinsic
variation.  In their classic book on Gravitational Lenses, \cite{s92}
point out that, "It is very difficult to separate the intrinsic
variability of a source from lensing effects; the only hope is to find
features in the light curve characteristic of microlensing
events".  This is largely the motivation for the present work.

In this paper, an additional expected consequence of dark matter made
up of stellar mass compact bodies is examined.  At a sufficiently high
redshift the optical depth to microlensing $\tau$ will be large enough for
caustic webs to form in the amplification pattern of the lenses.  This
will result in distinctive cusp-like features to be present in the
microlensing light curves.  The identification of such features provides
a necessary consistency check for the hypothesis that the dark matter is
made up of compact bodies.  It is the purpose of this paper to look for
such events.

The structure of the paper is as follows.  Section~\ref{dat}
describes the quasar monitoring programmes which were used to identify
candidate caustic crossing features, and computer simulations to produce
templates for identifications.  Section~\ref{met} gives examples of more
realistic computer simulations to be compared with observed features in
quasar light curves, and discusses the processes which produce them.
Section~\ref{res} shows candidate caustic crossing events and relates them
to simulated microlensing events, as well as giving some light curve
statistics relevant to microlensing.  The final sections are for
Discussion and Conclusions.

\section{Data}
\label{dat}

This section is devoted to a description of the quasar monitoring
programmes which form the basis for the search for microlensing events,
and a description of the microlensing simulations which have been used as
templates for identifying caustic crossing events.

\subsection{Observations of quasar light curves}
\label{obs}

Since the first observations of quasar optical variability \citep{m63},
there have been a number of multicolour monitoring programmes, pioneered
by \cite{g99} with a sample of 42 nearby quasars observed for 7 years.
This work was followed by a larger photographic survey \citep{h96,h07} in
which some 1000 quasars were monitored for a period of 26 years, with data
held in the SuperCOSMOS Science Archive(SSA)\footnote{http://ssa.roe.ac.uk}
and more recently the SDSS Legacy Survey \citep{m12} in Stripe 82\footnote
{https://faculty.washington.edu/ivezic/macleod/qso-dr7/Southern.html} with
light curves for nearly 10,000 quasars covering up to 8 years. These
surveys were designed to understand the nature of the observed variations
in quasar light, and form an excellent basis for the search for caustic
crossing events.

The first survey large enough to be useful for detecting microlensing
events was the photographic monitoring programme \citep{h96,h07}
undertaken with the UK 1.2 Schmidt telescope at Siding Springs Observatory,
Australia, in ESO/SERC Field 287 centred on ${\rm 21^h 28^m}, -45^\circ$
(1950).  The plates were digitised by the SuperCOSMOS measuring machine
at the Royal Observatory, Edinburgh \citep{h01} to give calibrated
magnitudes for the several hundred thousand images on the
6$^\circ$ x 6$^\circ$ plates.  In a recent update of the survey
\citep{h22}, the catalogue now contains some 1033 quasar light curves in
the $B_J$ (close to $g$) and $R$ passbands, with yearly mean magnitudes
for 26 years from 1977 to 2002 and photometric errors $< 0.04$
magnitudes.

\subsection{Microlensing simulations}
\label{mic}

\begin{figure}
\centering
\begin{picture} (150,410) (50,-10)
\includegraphics[width=0.49\textwidth]{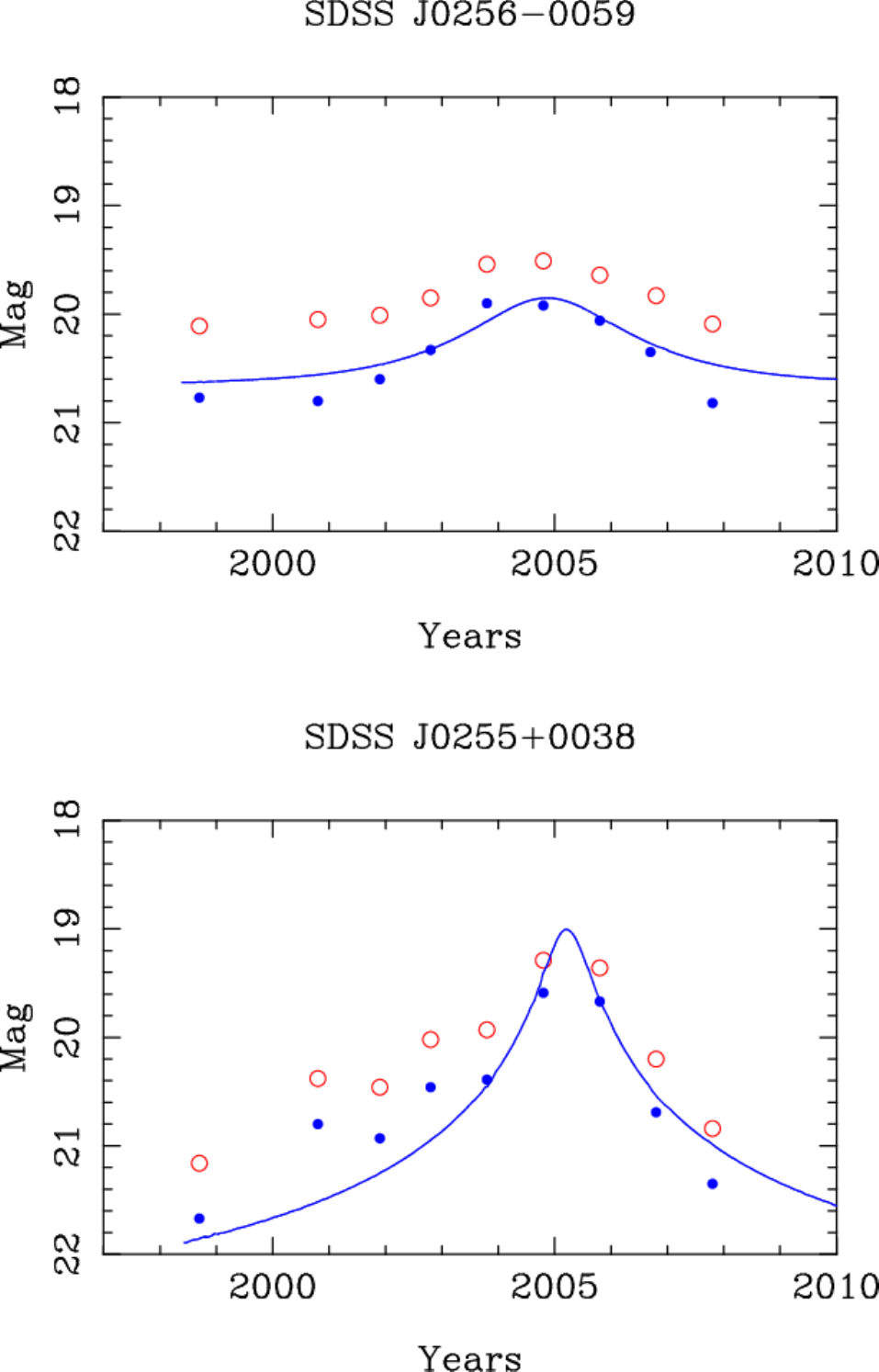}
\end{picture}
\caption{Examples of candidate microlensing events from Luo et al. (2020).
 Blue filled and red open circles show average yearly magnitudes for the
 $g$ and $r$ band filters respectively.  The continuous blue line shows a
 Paczy\'{n}ski profile (Alcock et al. 1996) fitted to the $g$-band data.}
\label{fig1}
\end{figure}
\begin{figure}
\centering
\begin{picture} (105,260) (40,-50)
\includegraphics[width=0.41\textwidth]{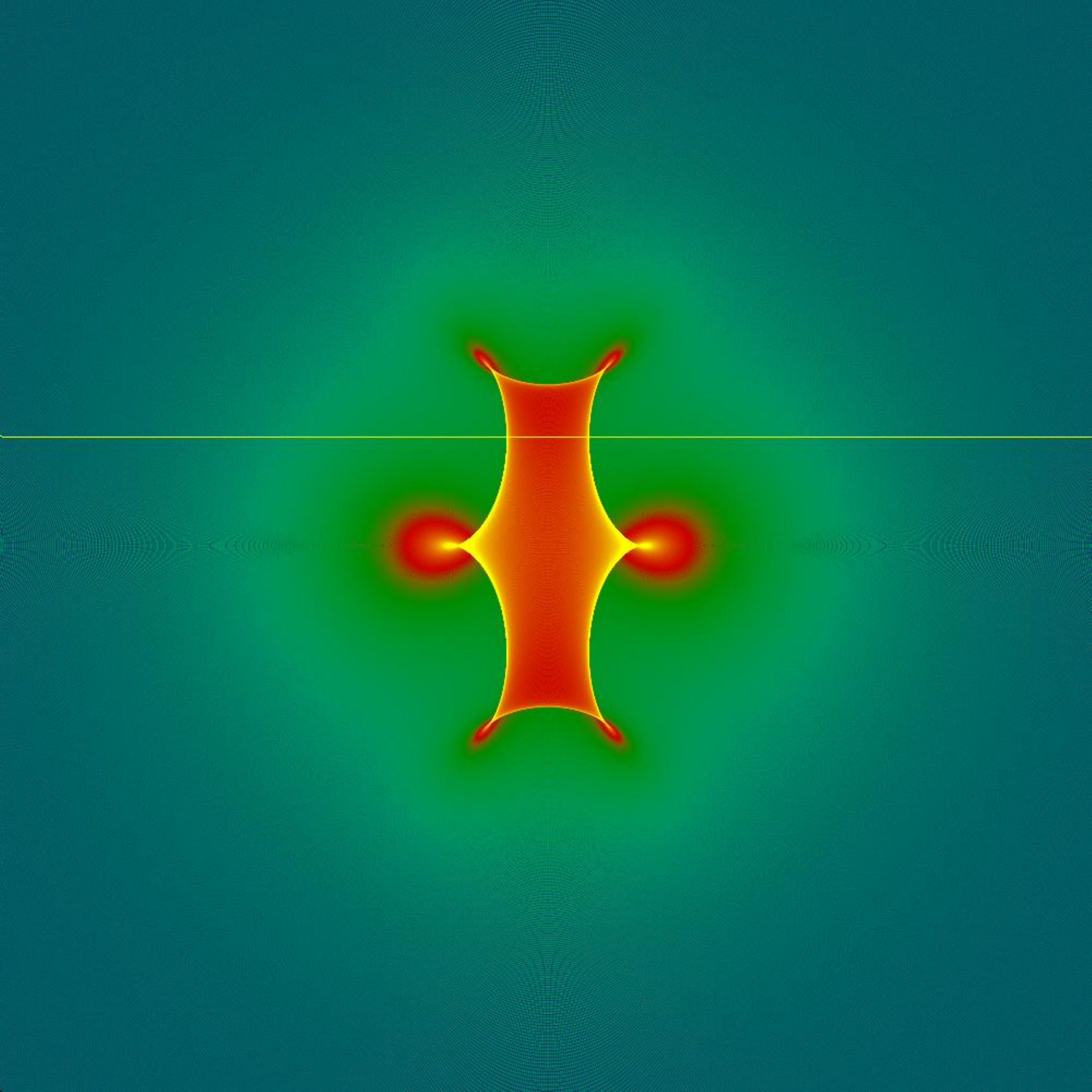}
\end{picture}
\begin{picture} (150,100) (45,-10)
\includegraphics[width=0.49\textwidth]{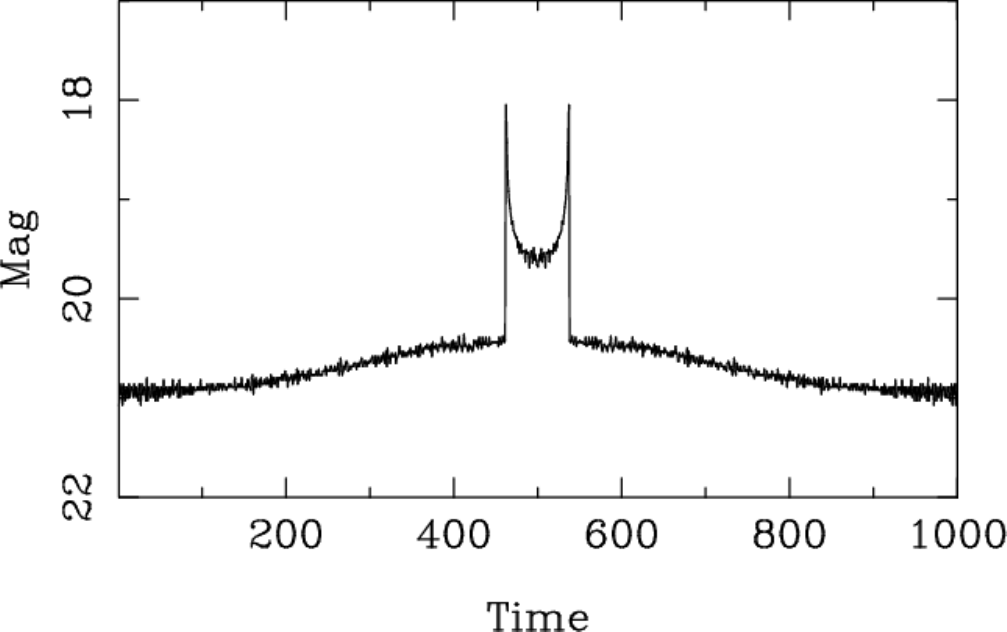}
\end{picture}
\caption{Microlensing magnification pattern for two equal mass lenses
 separated by $1.0 R_E$, in arbtrary units of time and magnitude.  The
 plot shows critical curves similar to those for the analytic solution
 illustrated in Figure 2 of Schneider \& Weiss (1986).  The bottom panel
 shows the light curve traced out by the yellow horizontal line, and
 indicates the magnitude scale.}
\label{fig2}
\end{figure}

\begin{figure*}
\centering
\begin{picture} (0,200) (240,-10)
\includegraphics[width=0.32\textwidth]{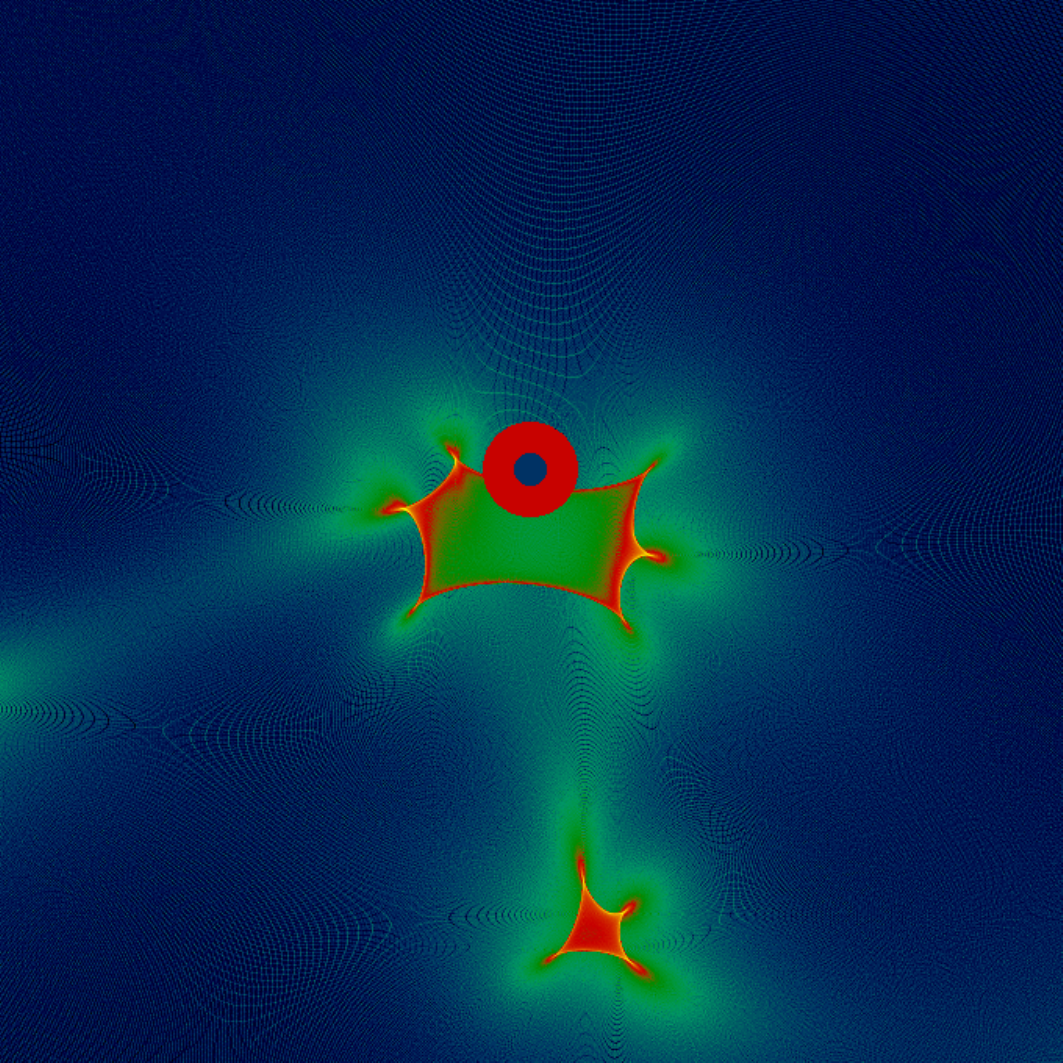}
\end{picture}
\begin{picture} (0,0) (70,-10)
\includegraphics[width=0.32\textwidth]{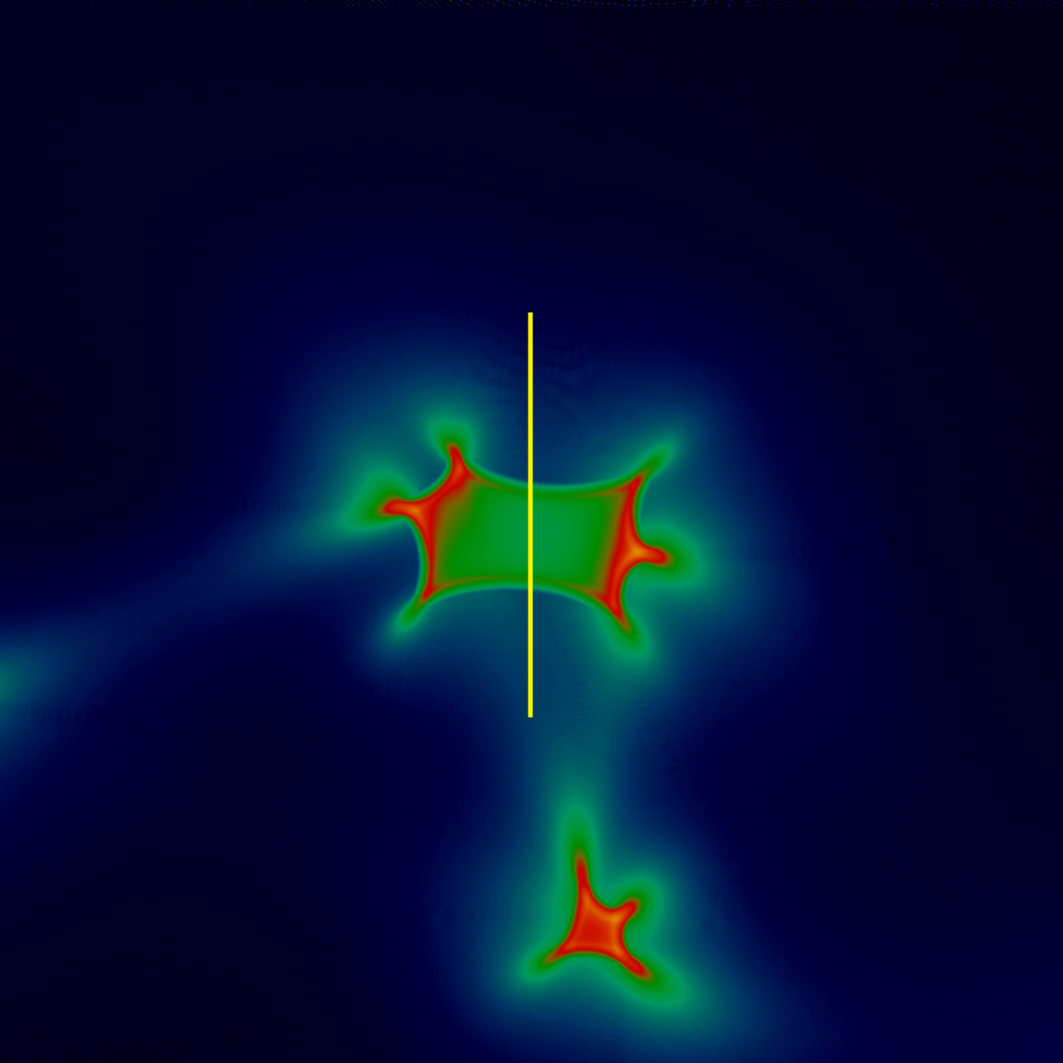}
\end{picture}
\begin{picture} (0,0) (-100,-10)
\includegraphics[width=0.32\textwidth]{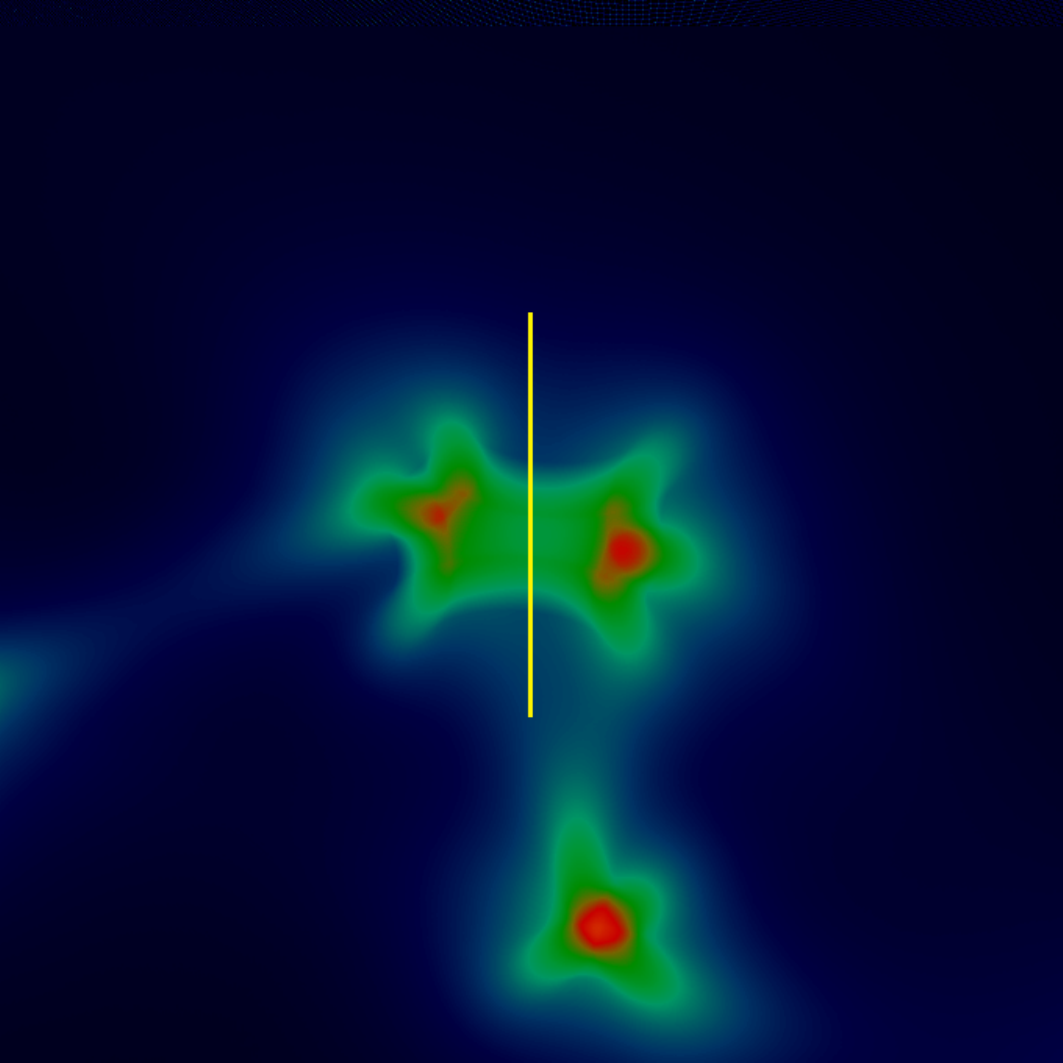}
\end{picture}
\caption{Microlensing magnification patterns in arbitrary units of
 magnitude for a population of 0.8 M$_\odot$ bodies with parameter values
 $\kappa_* = 0.1$, $\kappa_c = 0$, and $\gamma = 0$.  The frames consist
 of $680^2$ pixels and have a side length of approximately 5.5 Einstein
 radii.  The left hand panel is  for a point source, and the centre and
 right hand panels assume a uniform  disc for the source with radius 1 and
 3 lt-day respectively.  Assuming redshifts $z =0.5$ and $z =2.0$ for lens
 and source, this corresponds to radii 0.06 and 0.17 $R_E$ respectively.
 To give an idea of relative sizes, in the left hand panel discs with
 radii 1 and 3 lt-day are superimposed on the caustic pattern.  The
 yellow lines indicate the trajectory of a quasar across the amplification
 pattern during a caustic crossing event, and the resulting light curves
 are plotted in Fig.~\ref{fig4}, and indicate the magnitude scale.
 The zeropoint of the colour code is arbitrary.}
\label{fig3}
\end{figure*}

The unambiguous detection of microlensing events in quasar light curves
presents a number of challenges, mostly connected with the difficulty of
distinguishing between intrinsic variations in the quasar luminosity as
opposed to amplification of the accretion disc by microlensing events.
A number of factors will contribute to the likelihood of microlensing
events being detected, including the redshift of the quasar, the
crowdedness of the population of lenses, the dimensions and structure of
the quasar accretion disc, the size of the sample of quasars, and the
length of time for which the quasars are monitored.

The particular focus of this paper is to look for evidence of microlensing
amplification by a population of compact bodies, which must be
distinguished from changes in the luminosity of the quasar accretion disc.
For isolated lenses, microlensing amplification will take the form of a
characteristic bell-shaped profile \citep{p86,a96}, commonly referred to
as a Paczy\'{n}ski profile, with source amplification $A$ given by

\begin{equation}
A = \frac{u^2 + 2}{u \sqrt{u^2+4}}
\label{eqn1}
\end{equation}

\noindent where $u = b/R_E$ is the impact parameter, and $b$ and $R_E$
are the separation of the lens from the line of sight to the quasar and
the Einstein radus of the lens respectively.  However, when a source
is amplified by two lenses, the amplification patterns combine in a
non-linear way to produce astroid like features, well illustrated for the
case of a two-point-mass lens by \cite{s86}.  As the value of $\tau$
increases, complex magnification patterns dominated by astroid-like
caustics emerge.  The transition from the low optical depth regime is
estimated to occur for a value of $\tau$ of around 0.1 \citep{k97}, which
for dark matter in the form of compact bodies in a $\Lambda$CDM cosmology
implies a redshift for the quasar of $z \sim 1.3$ \citep{f92}.  It is
therefore to be expected that in a large sample of quasars with say
$z > 1.5$, occasional caustic crossing events may be observed in quasar
light curves.

In a recent paper \citep{l20} largely devoted to characterising the
optical light curves of extreme variability quasars, the authors also
identify a sample of 16 quasars with bell shaped light curves which they
present as candidate microlensing events.  The main database for their
search is the SDSS Legacy photometric monitoring programme in Stripe
82, with additional measures to extend the light curves from the
Pan-STARRS1 data archive\footnote {https://catalogs.mast.stsci.edu/
panstarrs/} and the Dark Energy Survey\footnote{https://des.ncsa.illinois
.edu/releases/dr2}.  Two examples of light curves from the list of
microlensing candidates identified by \cite{l20} are plotted in
Fig.~\ref{fig1}.  To clarify the trend of the data, the original
measurements have been averaged into yearly bins, and for comparison a
Paczy\'{n}ski profile has been fitted to the $g$-band data.

Although these quasar light curves are certainly plausible candidates for
microlensing events, the shape alone is not sufficient for an unambiguous
classification as a Paczy\'{n}ski profile from Eq.~\ref{eqn1}.
When two or more lenses lie close to the line of sight to a quasar, within
say one Einstein radius, the lenses combine in a non-linear way to produce
an astroid-like magnification pattern with clearly defined critical
curves.  The light curves resulting from the transit of a source across
such a complex magnification pattern are not in general susceptible to
analytic description, and must be modelled by computer simulation.  To
achieve this, the ray tracing software of \cite{w99} has been used to
simulate the non-linear magnification pattern resulting from multiple lens
configurations.  In these situations, the resulting light curves show
characteristic cusp-like features which are far more readily identified
as microlensing features than Paczy\'{n}ski profiles.

\section {Methods}
\label{met}

The first step up in complexity from the single lens Paczy\'{n}ski profile
is the case of two equal mass lenses, which has in fact been treated
analytically by \cite{s86}, who trace out the critical curves for a range
of lens separations.  As an example, the ray tracing software of
\cite{w99} has been used to simulate the magnification pattern for two
equal mass lenses separated by $1.0 R_E$, illustrated in the top panel of
Fig.~\ref{fig2}. The bottom panel shows the light curve corresponding to
the simulated amplification pattern where the crossing of the critical
curves can be seen as sharp cusps.  Examples of light curves where these
characeristic cusps are clearly defined are well illustrated by
simulations from \cite{r91}.  In a search for examples of microlensing
events, these cusp like features are far more distinctive and unambiguous
than the bell shape of a Paczy\'{n}ski profile as illustrated in
Fig.~\ref{fig1}.

\begin{figure}
\centering
\begin{picture} (150,200) (60,0)
\includegraphics[width=0.49\textwidth]{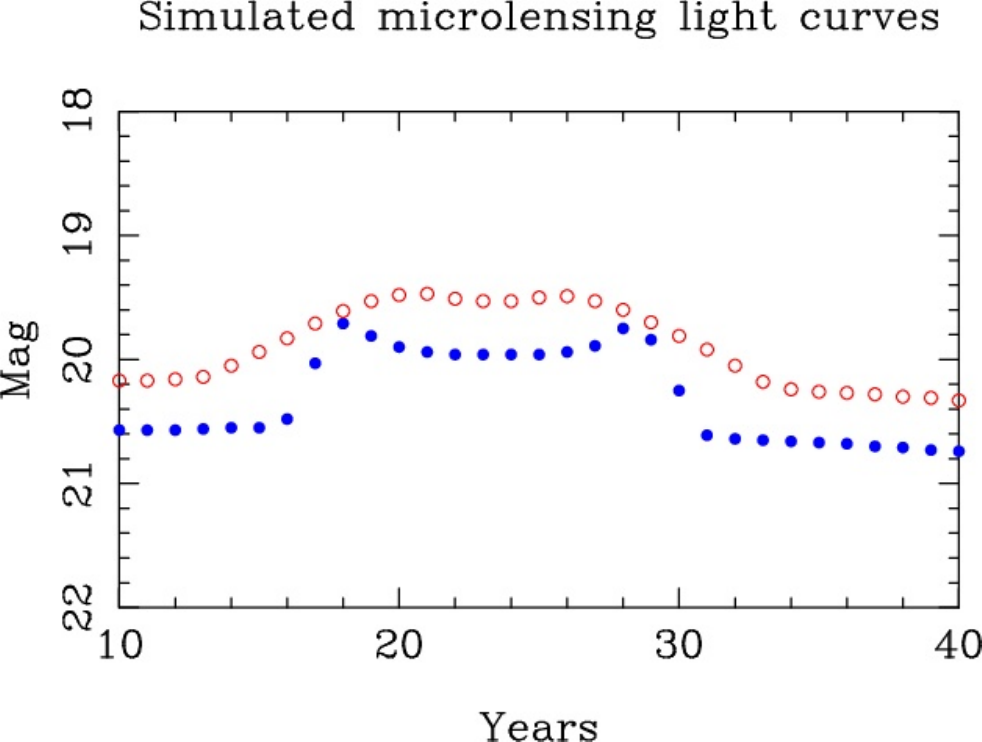}
\end{picture}
\caption{Simulated light curves corresponding to the yellow tracks in
 Fig.~\ref{fig3}.  The tracks have been plotted as yearly bins to
 facilitate comparison with observations.  The blue filled circles are
 for a source size of 1 lt-day and the red open circles for 3 lt-day.
 The magnitude scale has an arbitrary zeropoint.}
\label{fig4}
\end{figure}

The amplification pattern in Fig.~\ref{fig2} assumes a point source.  In
more realistic cosmological simulations \citep{h20a,h22}, the source size
in the form of the quasar accretion disc will not necessarily be
negligible compared with the Einstein radius of the lenses.  To illustrate
the effect of a finite source size, Fig.~\ref{fig3} shows a cut-out of
caustic features from a large area simulation with surface density
$\kappa_*$ in compact bodies or lenses, where $\kappa_* = 0.1$.  The
timescale for the simulation is set by assuming a stellar mass lens,
typical values for the source and lens redshifts, and the conventionally
adopted value of 600 km sec$^{-1}$ for the net transverse velocity
\citep{k86}, recently refined by \cite{n20} for specific quasar systems.
The three panels show amplification patterns for a point source, and for
source sizes 1 and 3 lt-day.  To give an idea of relative sizes, in the
left hand panel discs corresponding in radius to 1 and 3 lt-day are
superimposed on the caustic pattern.  In order to understand the effect
that increasing the source size will have on changes in the the observed
brightness of a quasar traversing the caustic along the superimposed
yellow track in Fig.~\ref{fig3}, the resulting light curves are plotted
in Fig.~\ref{fig4}. The data points show average yearly magnitudes to
facilitate comparison with observed light curves.  Closed and open circles
show light curves for 1 and 3 lt-day quasar accretion disc sizes
respectively.

An important feature of microlensing light curves is statistical
symmetry, which can be used as a way to distinguish microlensing from
intrinsic variations.  As \cite{s92} point out, "microlensing should yield
light curves which are {\it statistically} symmetric".  On a cosmological
scale, the assumption of the isotropy of the Universe combined with the
geometric nature of the lens equations with no time components implies
that microlensing light curves will have no arrow of time.  In other
words, no statistical test will be able to determine the direction of the
time coordinate.  In practice this means that for a large sample of
quasars, monitored over a sufficiently long period of time, then for any
specific change in yearly magnitude, the numbers of increase and decrease
in brightness will be equal.  This can be used as a way of testing the
consistency of microlensing as a significant component of quasar
variability.

\section{Results}
\label{res}

The search for evidence of microlensing in quasar light curves is greatly 
facilitated if the target events are caustic crossings as simulated in
Figs~\ref{fig2} and~\ref{fig4}.  In this case the expected features are
far more distinctive and recognizable than the bell-shaped Paczy\'{n}ski
profiles in Fig.~\ref{fig1}.  Apart from the characteristic cusp-like
features of caustic crossings, the radial structure of the quasar
accretion disc can provide another parameter for identifying microlensing
events \citep{v24}. A fundamental property of accretion discs is a radial
temperature gradient as the disc cools with distance from the central
black hole.  If significant emission from an area of the accretion disc
lies within the Einstein radius of a microlens, then the accretion disc
will act as a point source and any amplification due to microlensing will
be achromatic.  However, in the case where the accretion disc is
comparable in size with the Einstein radius of the lens, it will be
effectively `resolved', with the bluer light near the centre acting as a
more compact source than the more diffuse red light.  This will result in
an observable and characteristic difference between the blue and red
passband microlensing light curves.  This expected pattern is illustrated
in Fig.~\ref{fig4}, where observations through a blue filter would be
represented by the filled circles, and those through a red filter by the
open circles.  As one might expect, the relatively sharp cusps of the
caustic crossing in the blue are smoothed out in the red.  This provides
another discriminant in searching for caustic crossing candidates in
samples of quasar light curves, where such a distinctive pattern of
variation is unlikely to be produced at random.

\begin{figure}
\centering
\begin{picture} (150,200) (60,0)
\includegraphics[width=0.49\textwidth]{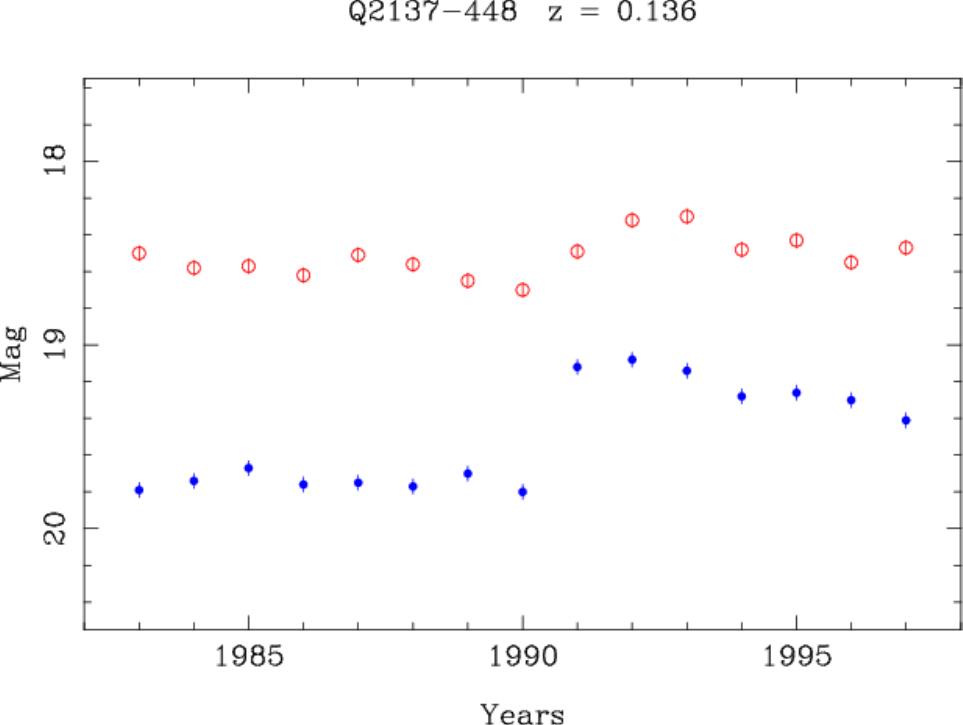}
\end{picture}
\caption{Light curve for a low redshift Seyfert galaxy from the Field 287
 survey.  Filled blue and open red circles show magnitudes for the $B_J$
 and $R$ passbands respectively.  Photometric errors as derived from the
 data are also plotted.}
\label{fig5}
\end{figure}

\begin{figure}
\centering
\begin{picture} (150,200) (60,0)
\includegraphics[width=0.49\textwidth]{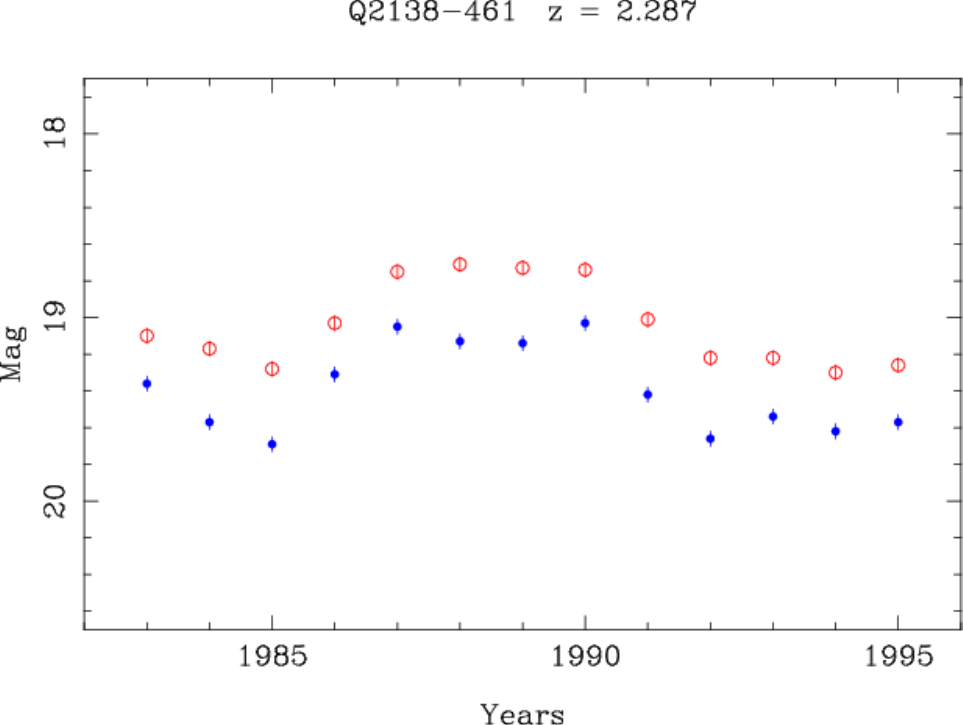}
\end{picture}
\caption{Light curve from the Field 287 survey showing a candidate caustic
 crossing event similar to the simulation in Fig.~\ref{fig4}.  Symbols are
 as defined for Fig.~\ref{fig5}.}
\label{fig6}
\end{figure}

\begin{figure*}
\centering
\begin{picture} (0,210) (250,-10)
\includegraphics[width=0.49\textwidth]{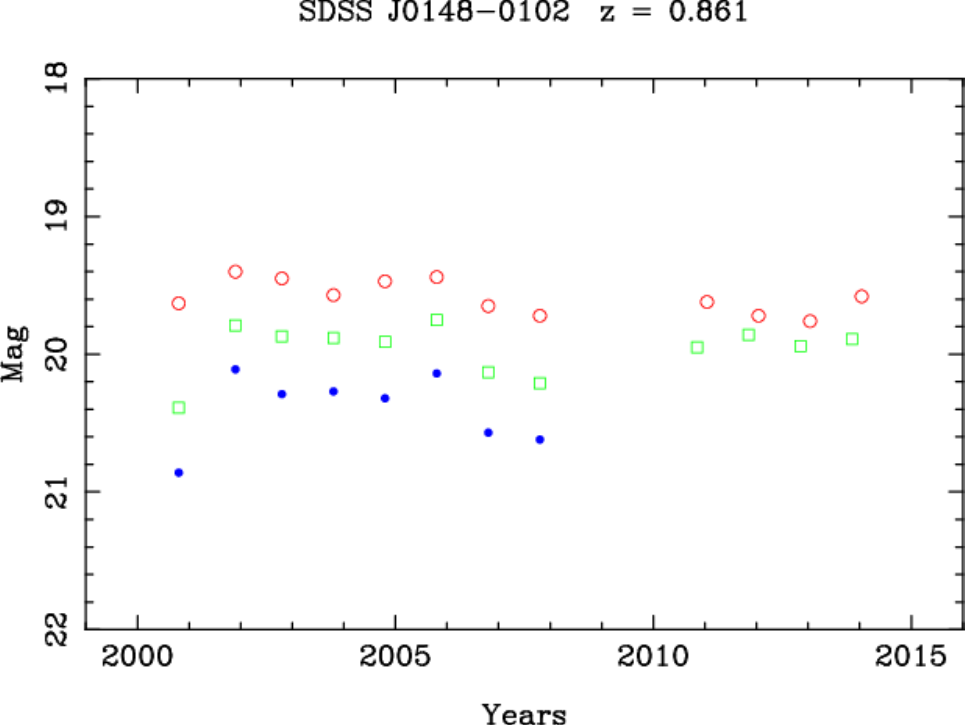}
\end{picture}
\begin{picture} (0,0) (-15,-10)
\includegraphics[width=0.49\textwidth]{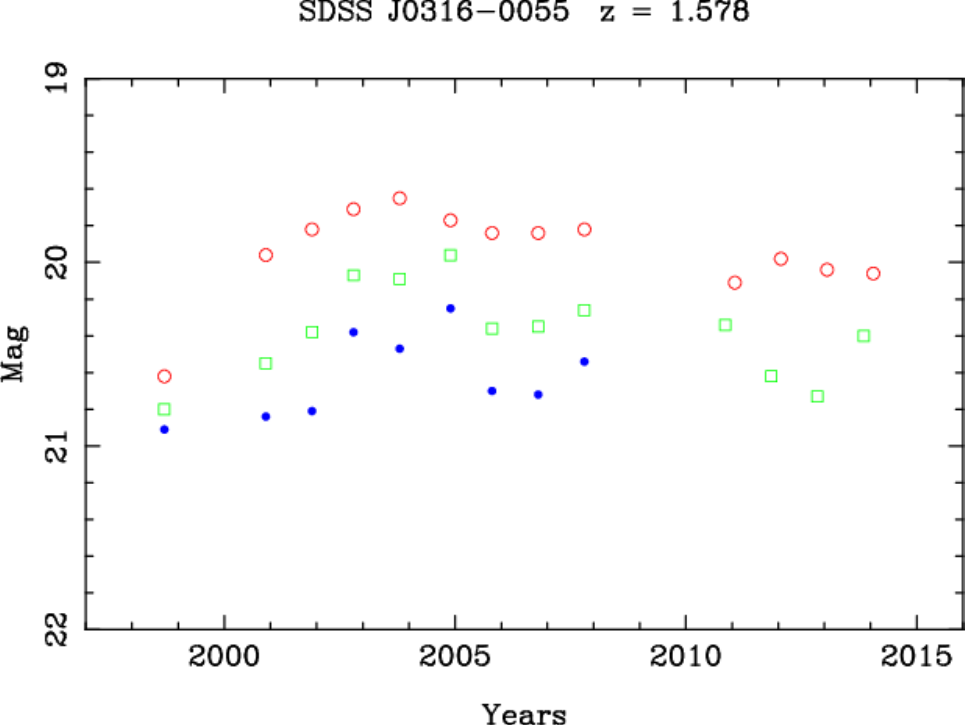}
\end{picture}
\caption{Quasar light curves from the SDSS Stripe 82 sample, showing
 candidate caustic crossing events.  Data for the years 1998-2008 are from
 the SDSS Stripe 82 data archive, and for 2010-2014 from the Pan-STARRS1
 data archive.  Filled blue circles, open green squares and open red
 circles show magnitudes for the $u$, $g$ and $z$ passbands respectively.
 The photometric error bars are smaller than the symbols.}
\label{fig7}
\end{figure*}

\begin{figure*}
\centering
\begin{picture} (300,280) (25,0)
\includegraphics[width=0.49\textwidth]{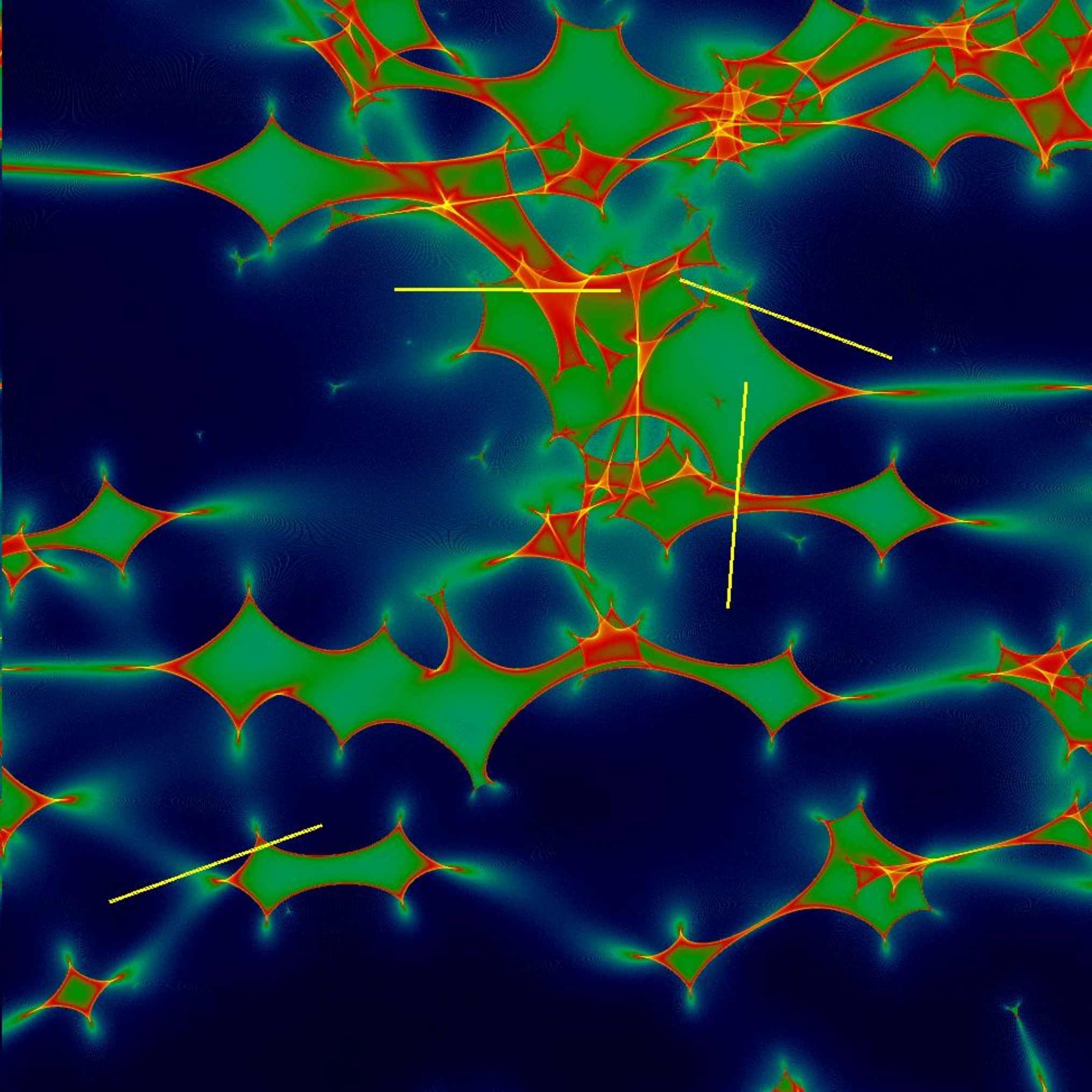}
\end{picture}
\begin{picture} (150,100) (70,0)
\includegraphics[width=0.49\textwidth]{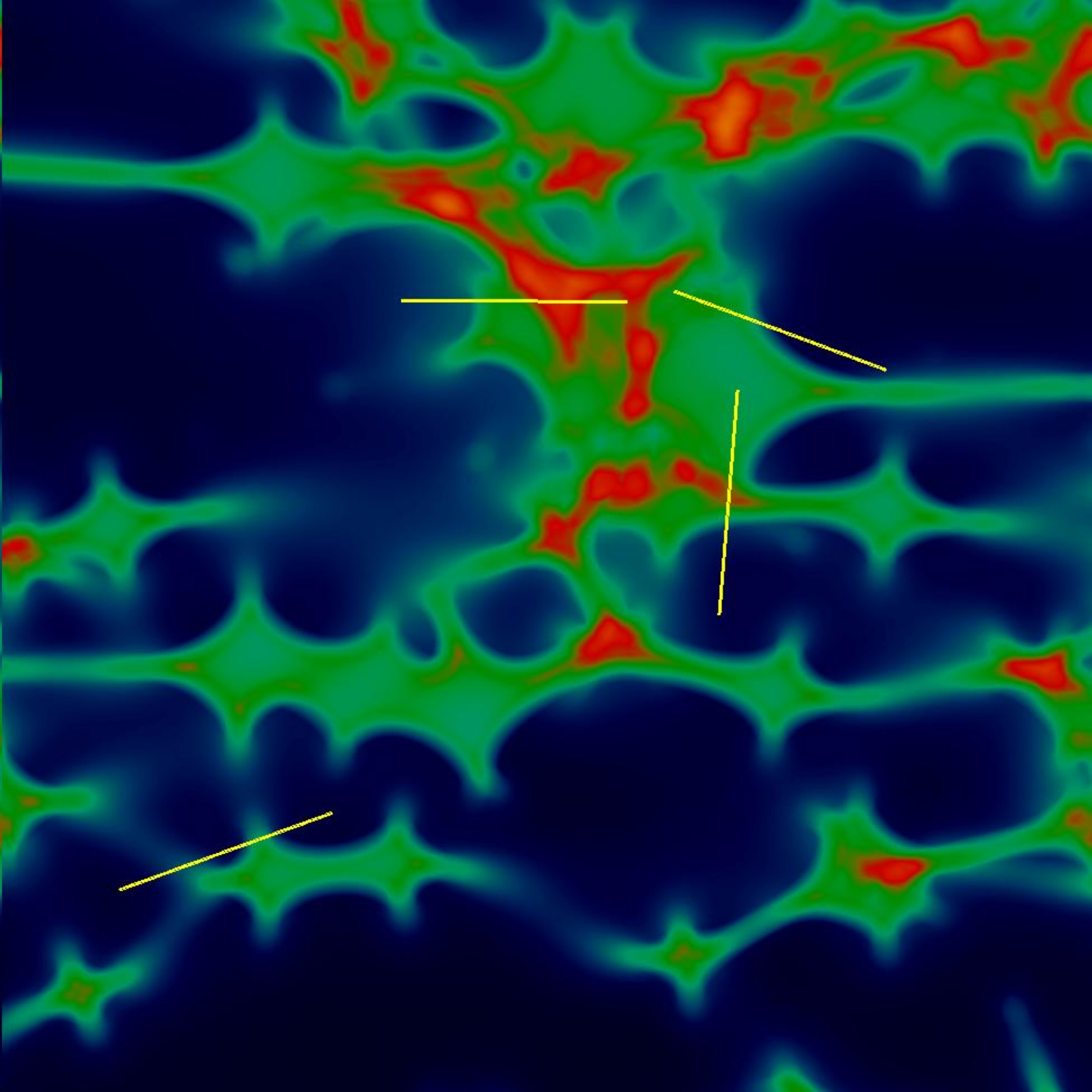}
\end{picture}
\caption{Microlensing magnification patterns for a population of 0.2
 M$_\odot$ bodies with parameter values $\kappa_* = 0.4$, $\kappa_c = 0$,
 $\gamma = 0.4$.  The frames consist of $1000^2$ pixels and have a side
 length of 8 Einstein radii.  The left hand panel is for a point source,
 and the right hand panel for a source size 2 lt-day, corresponding to
 0.23 $R_E$.  The yellow lines represent random 20 year tracks across the
 caustic pattern.}
\label{fig8}
\end{figure*}

\begin{figure*}
\centering
\begin{picture} (400,400) (55,0)
\includegraphics[width=0.99\textwidth]{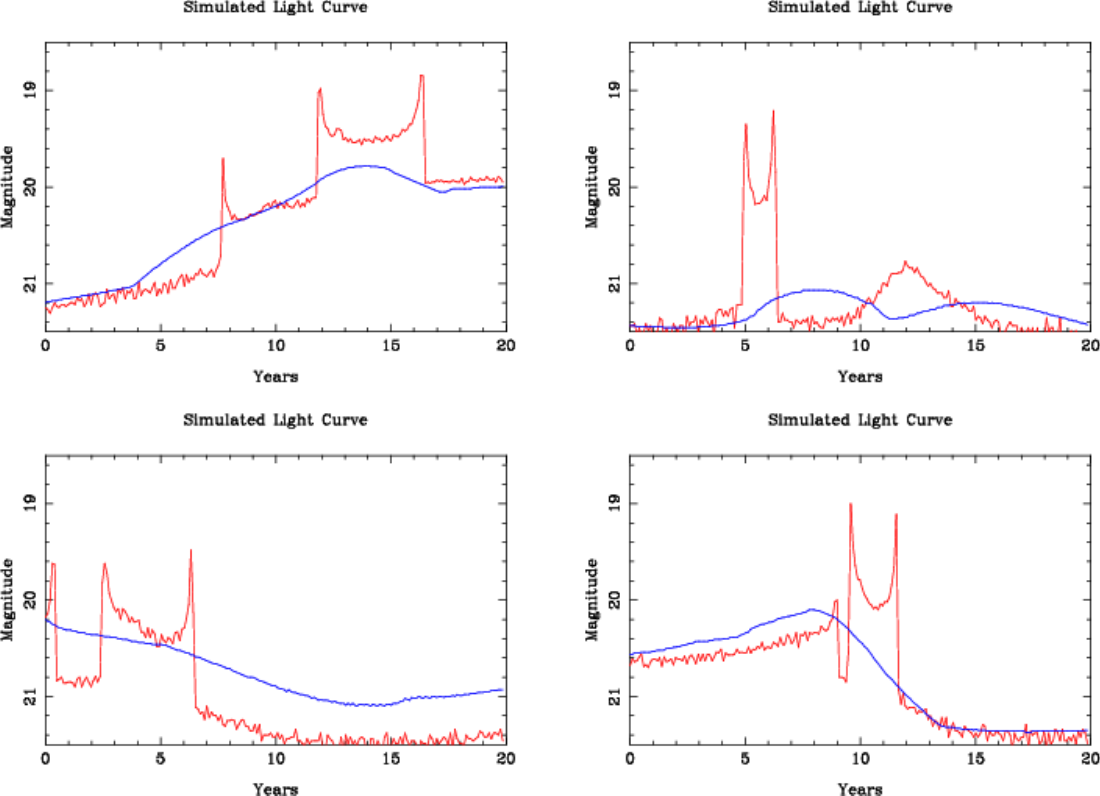}
\end{picture}
\caption{Simulated light curves corresponding to the yellow tracks in
 Fig.~\ref{fig8}.  The red lines are for a point source (left hand panel
 Fig.~\ref{fig8}), and the blue lines are for a source size of 2 lt-day
 (right hand panel Fig.~\ref{fig8}).}
\label{fig9}
\end{figure*}

The idea behind this paper is to look for evidence of microlensing
amplification as opposed to changes in the luminosity of the quasar
accretion disc.  In the event that there is a significant temperature
gradient across the accretion disc, from a hot blue core in the centre to
cool red outer parts, changes in luminosity typically manifest themseves
as flares originating in the blue core and propagating outwards to the
redder parts of the outer disc, which responds later and less violently
than the blue core.  This is illustrated in Fig.~\ref{fig5} which shows
blue and red passband light curves from the Field 287 survey for a Seyfert
galaxy, illustrating the expected features for a brightening of the
nuclear accretion disc.  The rapid brightening of the blue light which
dominates the hot centre is accompanied by a more gradual increase in red
light. However, a typical caustic crossing microlensing event has a
completely different light curve structure.  This is illustrated in
Fig.~\ref{fig6}, which shows another light curve from the Field 287
survey.  In this case the light curve bears a strong resemblance to the
caustic crossing simulation in Fig.~\ref{fig4}.  The overall structure is
symmetrical in time, with the blue passband showing cusp-like features
which are smoothed out in the red.  As discussed in relation to the
simulation, this is the result of the compact blue centre of the accretion
disc acting as a point source, whereas the more extended redder region is
effectively resolved by the microlens.  Features similar to those seen in
Fig.~\ref{fig6} regularly occur in samples of quasar light curves.  The
Stripe 82 sample from the SDSS Legacy catalogue of quasar light curves
contains many examples, two of which are shown in Fig.~\ref{fig7}.
Although the addition of extra photometric passbands to the light curves
adds some useful definition to the nature of the variations, the shortness
of the monitoring period of around 8 years makes it difficult to identify
features unambiguously. This is somewhat mitigated by extra epochs from
the Pan-STARRS1 data archive, but the inevitable gaps in the observations
make for difficulties in identifying features in the light curves.

In general, caustic structures are more complex than those
associated with the binary lens in Fig.~\ref{fig2}, as may be seen in
microlensing simulations with a higher surface density of lenses
$\kappa_*$, possibly amplified by a smooth distribution of matter
$\kappa_c$.  A typical example is shown in Fig.~\ref{fig8}, from a
simulation using the microlensing code of \cite{w99}.  The left hand panel
shows an amplification pattern for a distribution of point sources with an
optical depth to microlensing $\tau \approx 0.4$.  The right hand panel
shows the effect of increasing the source size, or in a cosmological
situation the effective radius of the quasar accretion disc.  The effect
is to blur out the sharp features associated with a point source.  To give
an idea of how this might affect the structure of quasar light curves, 4
yellow lines have been superimposed on the simulation, representing random
tracks of 20 years duration across the amplification pattern, as might be
observed for the light curve of a quasar in a cosmological setting. The 4
light curves corresponding to the yellow tracks are plotted in
Fig.~\ref{fig9}, where the effect of increasing the source size is clearly
illustrated.

\begin{figure}
\centering
\begin{picture} (150,400) (60,0)
\includegraphics[width=0.49\textwidth]{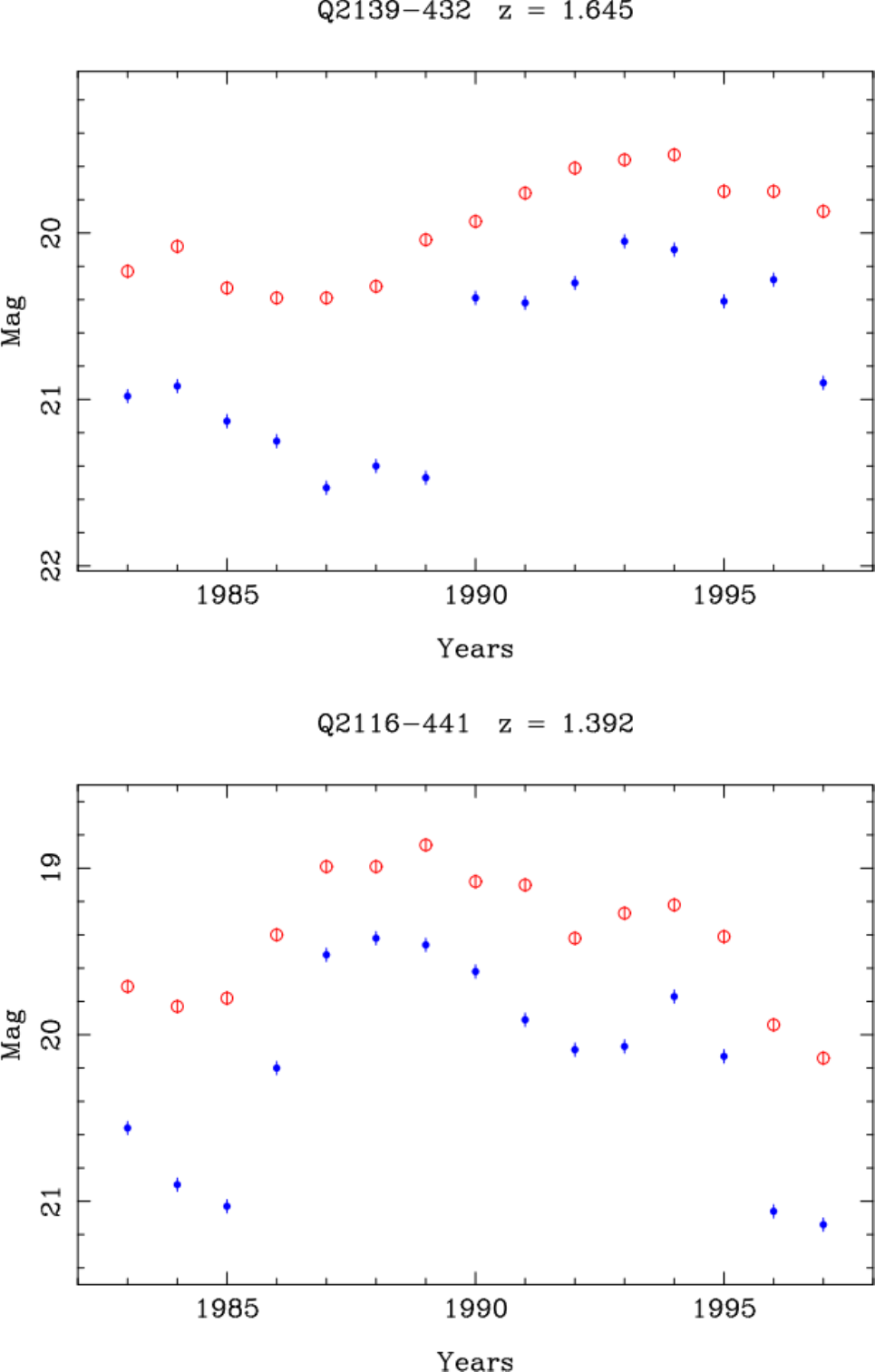}
\end{picture}
\caption{Two light curves from the Field 287 survey showing all the
 expected features of a complex caustic crossing.  Symbols are as defined
 for Fig.~\ref{fig5}.}
\label{fig10}
\end{figure}

\begin{figure}
\centering
\begin{picture} (150,200) (60,0)
\includegraphics[width=0.49\textwidth]{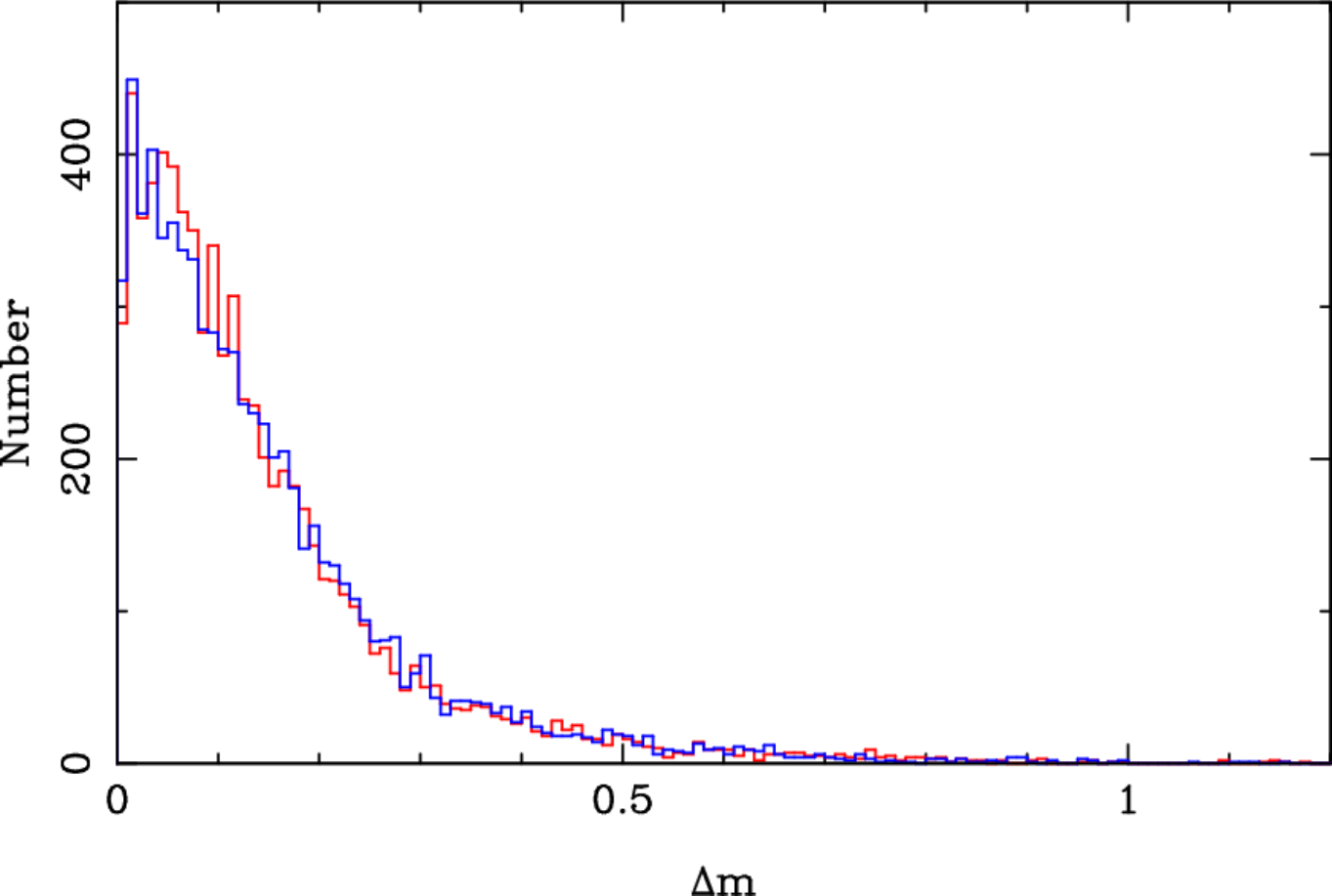}
\end{picture}
\caption{Histograms of yearly changes in magnitude in bins of 0.01 mag
 for light curves from computer simululations of microlensing
 amplifications.  Blue and red histograms are for brightening and
 faintening changes in magnitude respectively.}
\label{fig11}
\end {figure}

\begin{figure}
\centering
\begin{picture} (150,200) (60,0)
\includegraphics[width=0.49\textwidth]{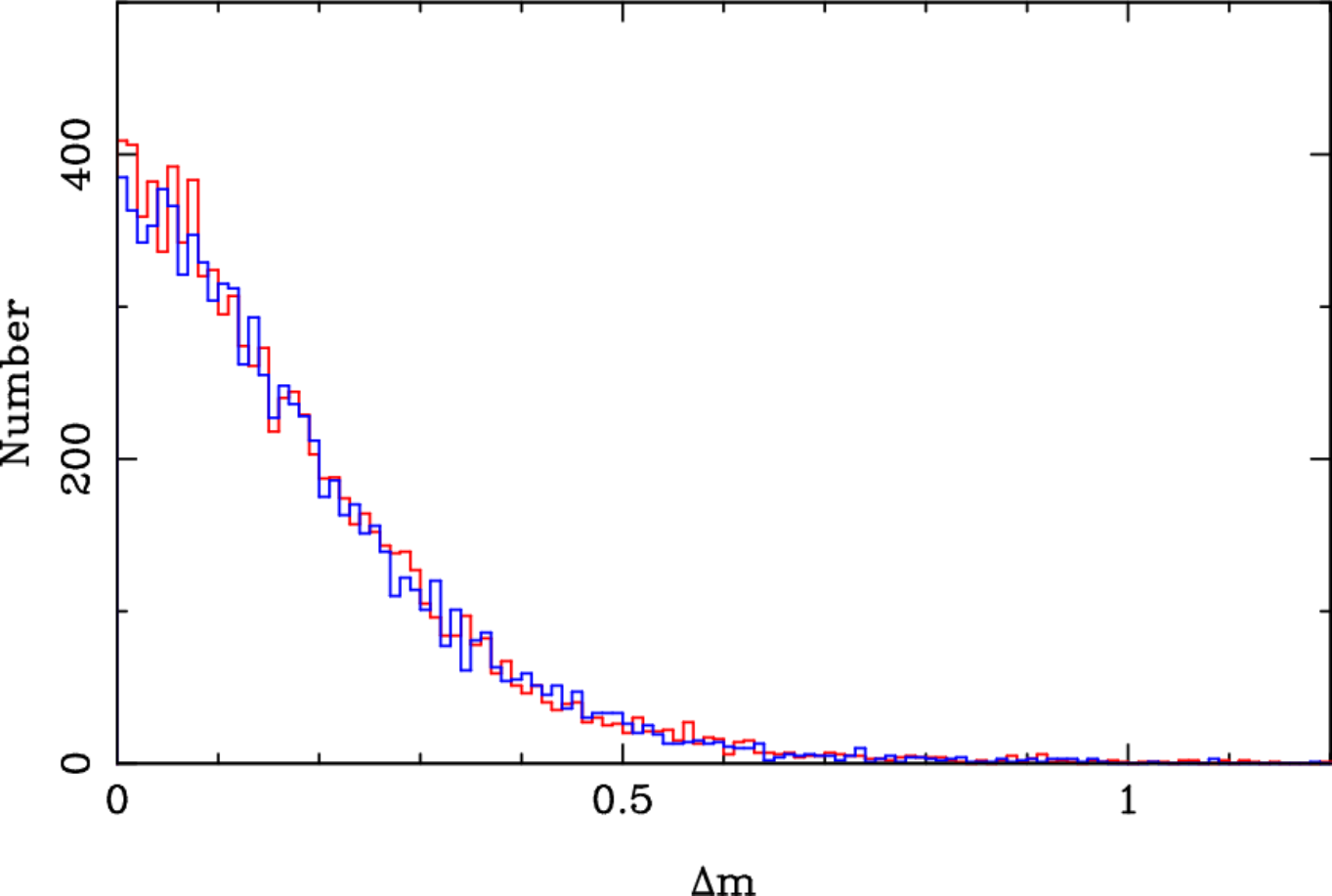}
\end{picture}
\begin{picture} (150,200) (60,0)
\includegraphics[width=0.49\textwidth]{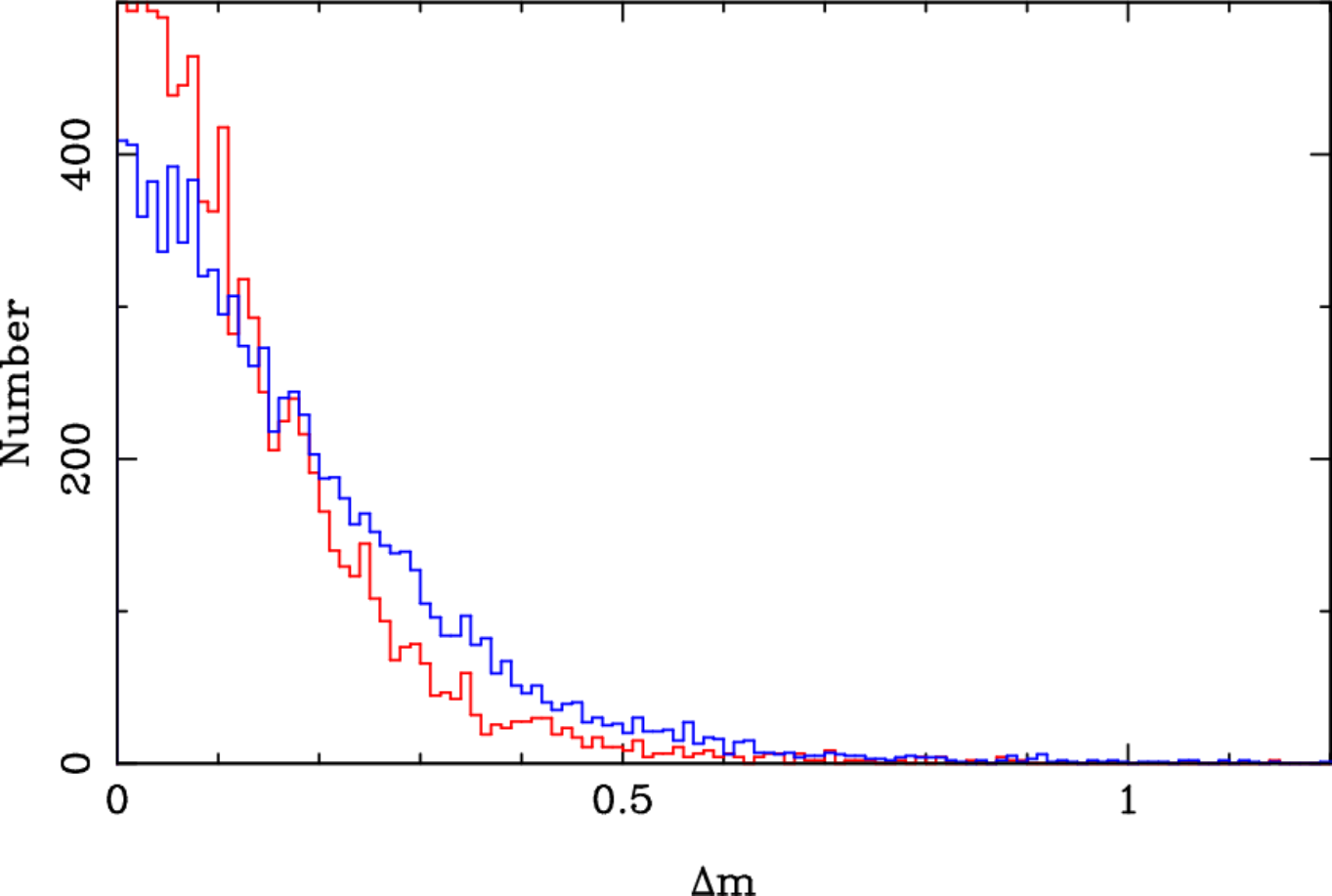}
\end{picture}
\caption{Histograms of yearly changes in magnitude in bins of
 0.01 mag for the Field 287 quasar sample.  The top panel shows
 brightening (blue histogram) compared with faintening (red histogram)
 changes for the blue passband light curves.  The bottom panel shows
 brightening in  the blue passband (blue histogram) compared with
 brightening in the red passband (red histogram).}
\label{fig12}
\end{figure}

In contrast to the light curve for a binary lens as illustrated in
Fig.~\ref{fig2} which is well-defined mathematically, or a simple caustic
crossing as simulated in Fig.~\ref{fig4}, there is clearly a random
element to the light curve of a quasar traversing the amplification
pattern in Fig.~\ref{fig8}.  Nonetheless, there are a number of
distinctive features in microlensing light curves which are hard to
explain as intrinsic variations in luminosity.  Perhaps the most
characteristic feature of a microlensing amplification of a quasar
accretion disc source with a radial colour gradient is the initial
brightening in the red light curve as the red outer part of the accretion
disc becomes amplified.  This is then followed by a sharper increase in
blue light and as the complex structure of the caustic is crossed, the
features are seen to be smoothed out in red light.  As the source leaves
the caustic the process is reversed, with the blue light declining more
quickly than the red, which persists for longer as the outer red part of
the accretion disc is finally crossed.  This is well illustrated in two
quasar light curves from the Field 287 sample illustrated in
Fig.~\ref{fig10}, which show all the features of a complex caustic
crossing.

The structure of the light curves in Fig.~\ref{fig10} raises another
interesting question.  Although they are clearly not symmetrical in time,
from a statistical perspective microlensing amplification has no arrow of
time.  It is thus to be expected that for any particular photometric
passband the timescales of increasing and decreasing light will be the
same.  This question of statistical symmetry which has been discussed
in general terms in Section~\ref{met} can be addressed by comparing
histograms of yearly increases and decreases in magnitude.  The
expectation for the structure of these histograms of magnitude differences
in quasar light curves on the basis that the variations are the result of
microlensing can be addressed by computer simulations.  The results are
illustrated in Fig.~\ref{fig11} which is based on a sample of light curves
from microlensing simulations similar to those illustrated in
Fig.~\ref{fig8}.  Here, the blue line is a histogram of yearly magnitude
increases, and the red line shows a similar histogram for yearly decreases
in magnitude.  It can be seen that there is no significant difference
between the two histogram, as expected from the geometrical origin of the
variations, and no statistical difference between the distribution of
increasing and decreasing variations.  The result of applying this test to
the light curves of the Field 287 sample of quasars is shown in the top
panel of Fig.~\ref{fig12}.  Again, there is no significant difference
between the two histograms, consistent with the expectations for
microlensing amplifications.  Similar histograms for red light show the
same effect, although histograms for red and blue light do not have the
same structure, as illustrated in the bottom panel of Fig.~\ref{fig12}.

The statistical symmetry of the quasar light curves as illustrated in
Fig.~\ref{fig12} can be used as a further consistency check for
the origin of the variations in quasar light.  In fact, symmetry in quasar
light curves has received little attention in the literature.  The issue
is discussed by \cite{k98} in the context of disc instability models of
quasar variation where they point out that in these models a slow rise in
luminosity is followed by a rapid decline.  They contrast this to a
starburst model where a central disturbance propagates through the disc
resulting in a rapid rise in luminosity, which is then followed by a
more gradual decline, as might be seen in a supernova for example.  The
symmetry illustrated in the top panel of Fig.~\ref{fig12} implies that the
observed quasar variations are not associated with models of this type,
although it is possible that the two modes of of variation could
combine to give the appearance of symmetry.  In fact it appears that
microlensing is the only well-studied physically motivated model of quasar
variation which predicts statistical symmetry in the light curves.
It is however important to emphasise that the observed statistical
symmetry in no way demonstrates that the quasar variations are the result
of microlensing.  It is perfectly possible that some as yet unidentified
mode of variation in the accretion disc is responsible for the observed
variability.  The significance for the results of this paper is just that
the symmetry observed in Fig.~\ref{fig12} is consistent with the
predictions of microlensing.

\begin{figure*}
\centering
\begin{picture} (0,420) (250,-220)
\includegraphics[width=0.49\textwidth]{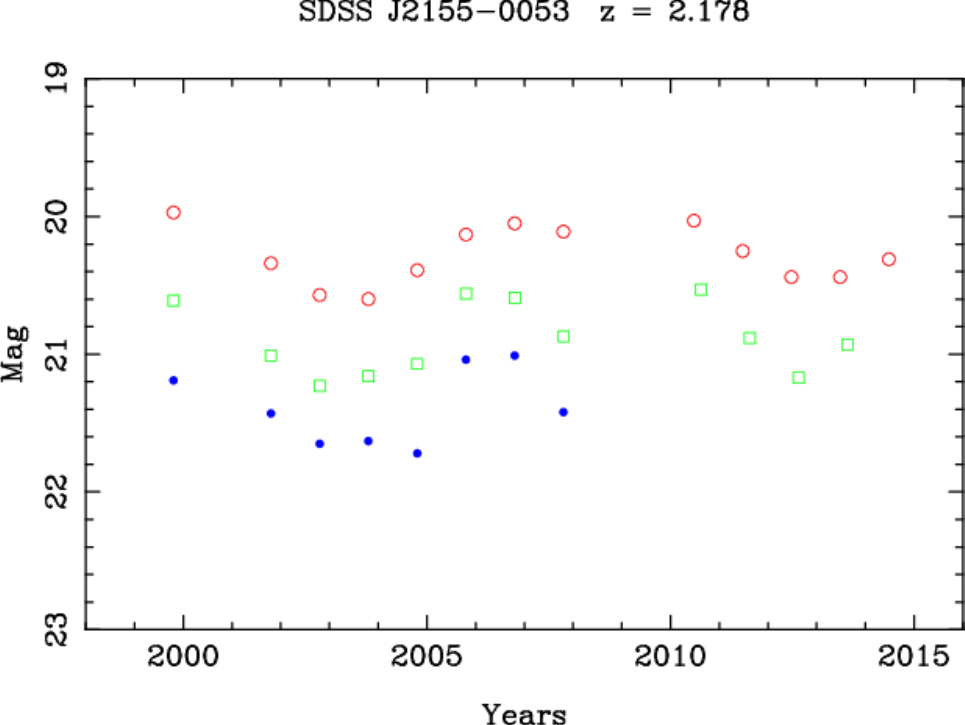}
\end{picture}
\begin{picture} (0,0) (-10,-220)
\includegraphics[width=0.49\textwidth]{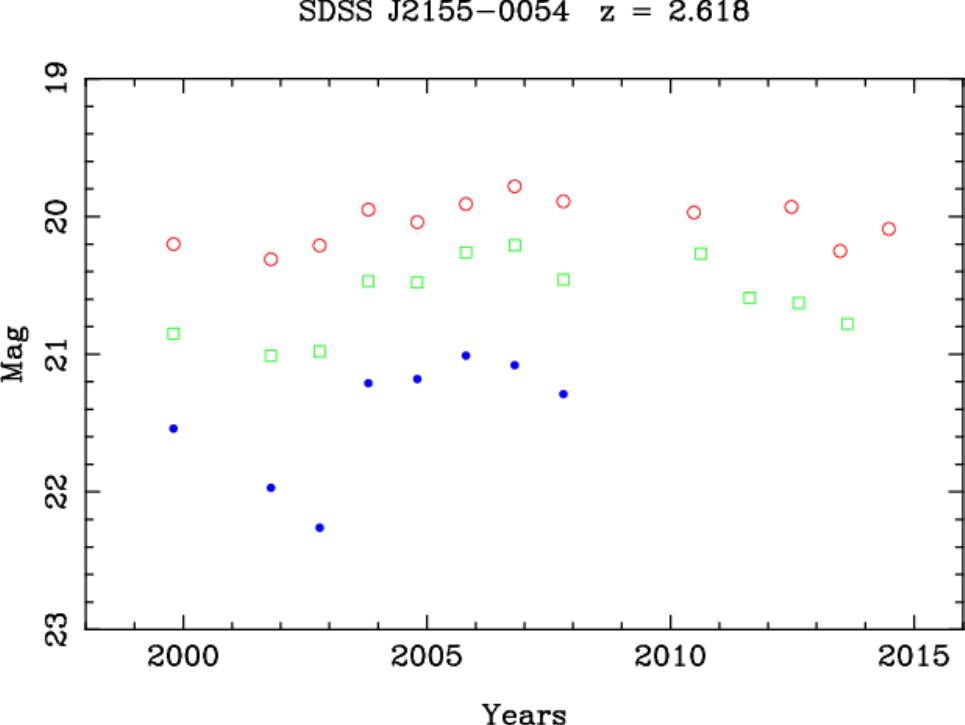}
\end{picture}
\begin{picture} (0,0) (250,-10)
\includegraphics[width=0.49\textwidth]{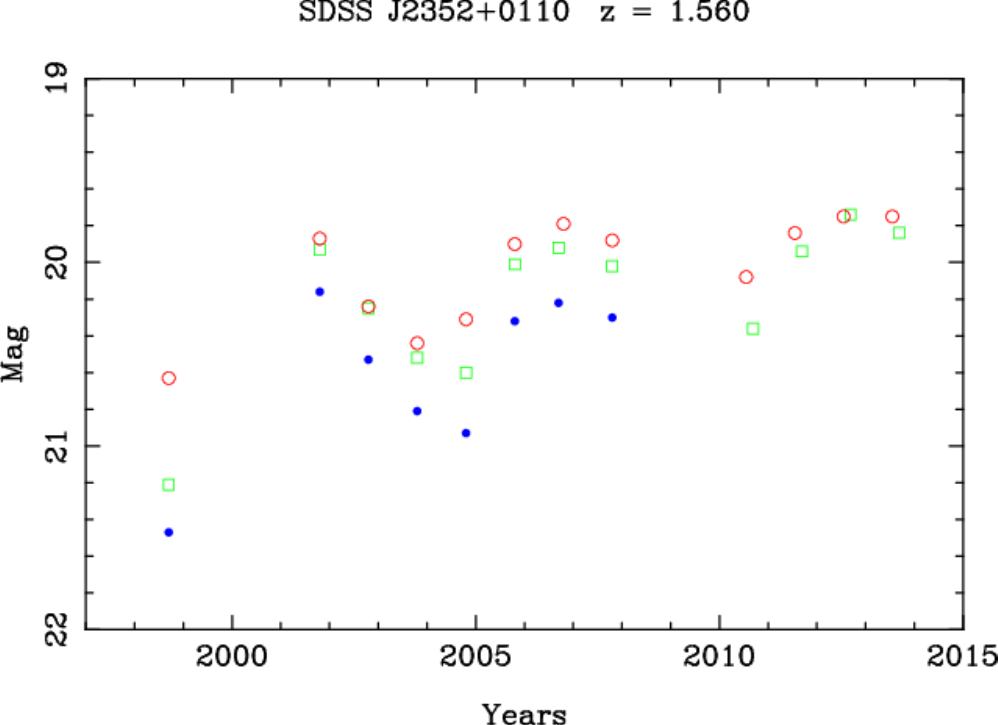}
\end{picture}
\begin{picture} (0,0) (-10,-10)
\includegraphics[width=0.49\textwidth]{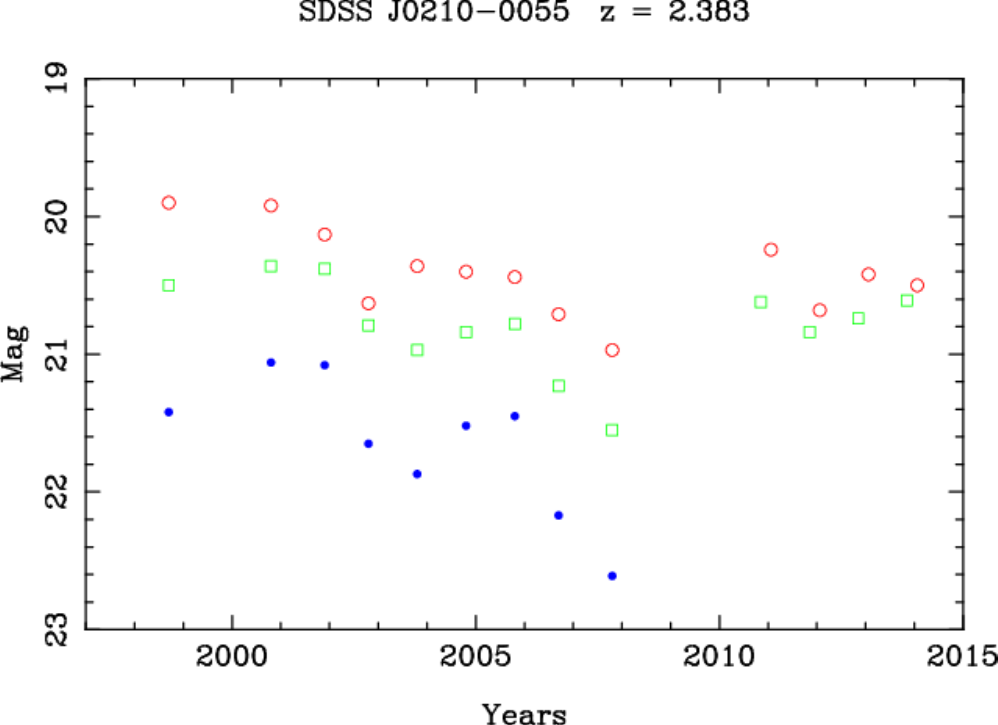}
\end{picture}
\caption{Quasar light curves from the SDSS Stripe 82 and Pan-STARRS1 data
 archives illustrating initial brightening in the red.  Symbols are as
 for Fig.~\ref{fig7}.}
\label{fig13}
\end{figure*}

The characteristic initial gentle increase in red light followed by a
sharper increase in blue light can form a useful diagnostic for
microlensing amplifications of an accretion disc with a temperature or
colour gradient.  The sample of quasar light curves in Stripe 82 provides
a source for the identification of this mode of variation, and
Fig.~\ref{fig13} illustrates four examples.  Although the structure of
the light curves can be quite complex, all four show an early slow
increase in red light followed by a more rapid and larger rise in blue
light.  The short time coverage and uneven sampling of the Stripe 82 light
curves unfortuneately obscures the full range of the caustic crossings.

\section{Discussion}
\label{dis}

\begin{figure}
\centering
\begin{picture} (150,200) (60,0)
\includegraphics[width=0.49\textwidth]{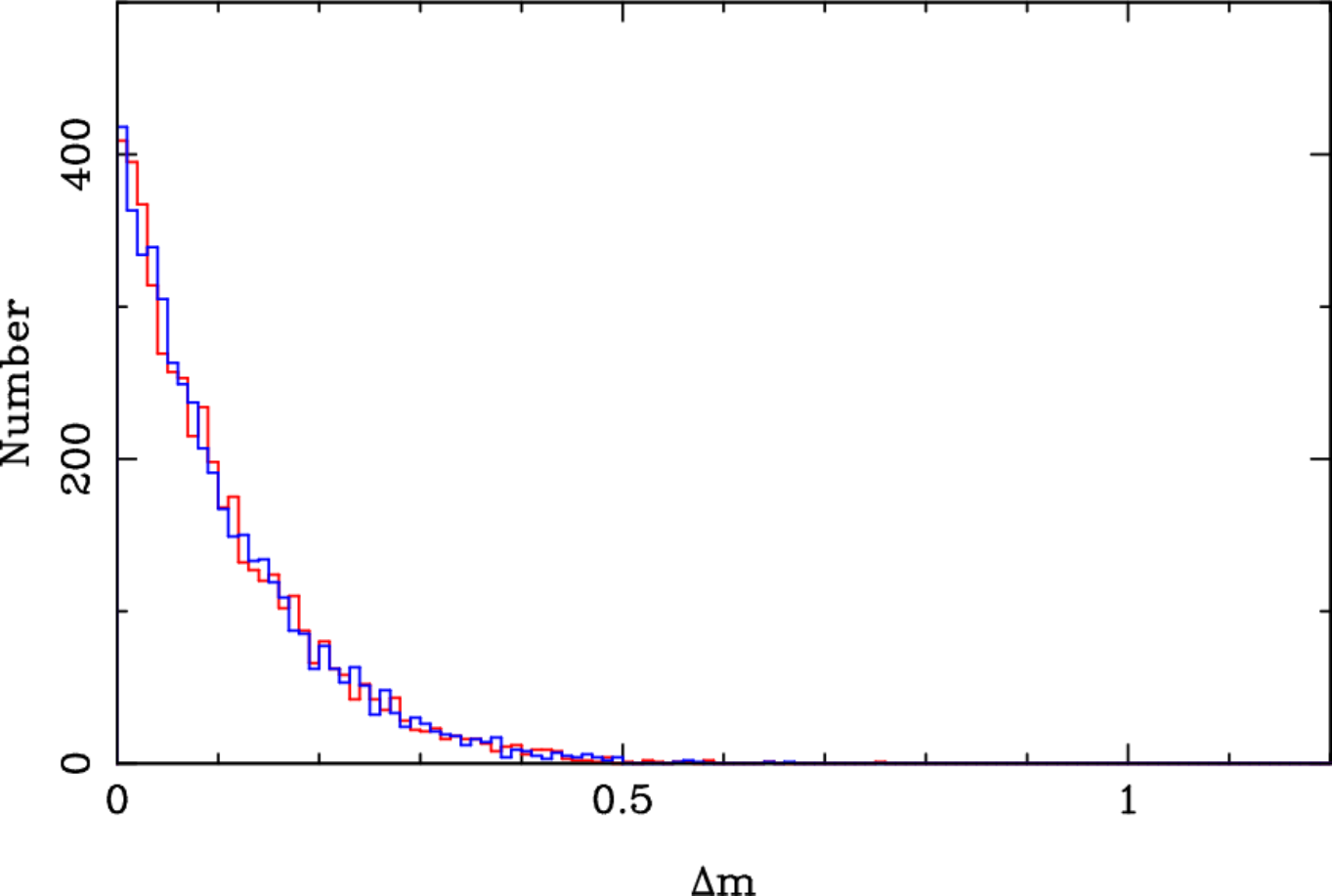}
\end{picture}
\caption{Histograms of yearly changes in magnitude in bins of 0.01 mag
 for light curves from computer simulations of variations from a damped
 random walk process.  Blue and red histograms are for brightening and
 faintening changes in magnitude respectively.}
\label{fig14}
\end {figure}

\begin{figure}
\centering
\begin{picture} (150,200) (60,0)
\includegraphics[width=0.49\textwidth]{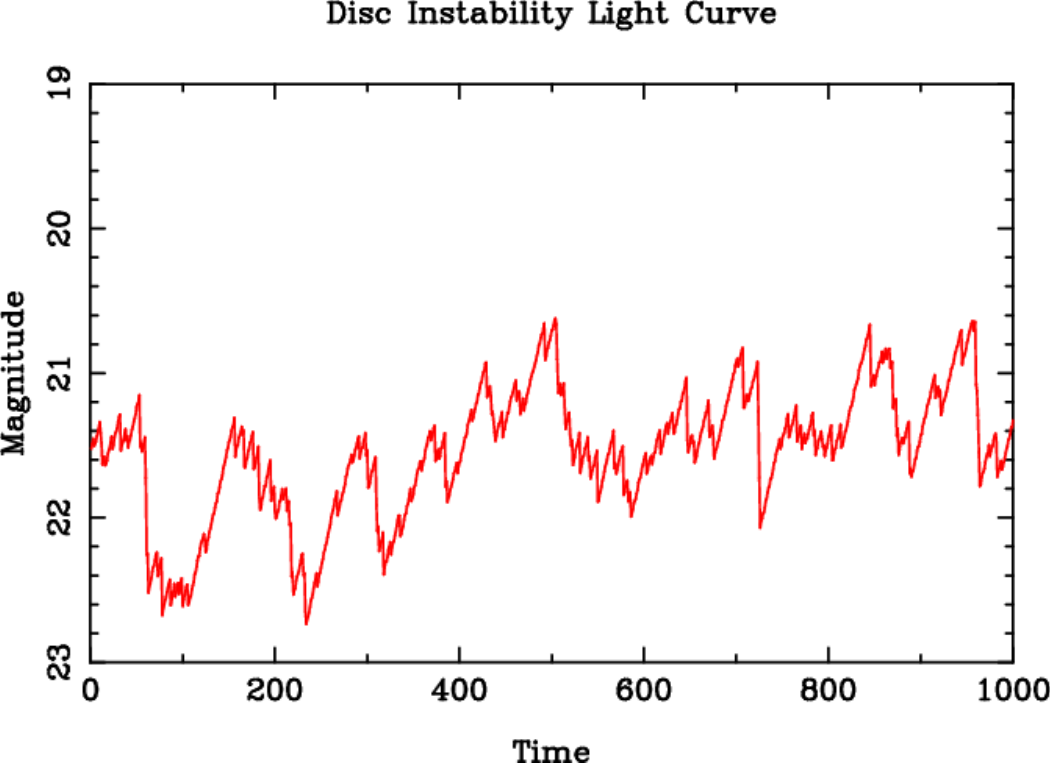}
\end{picture}
\caption{An example of a light curve generated by the self organizing
accretion disc model, with arbirary units of time and magnitude.}
\label{fig15}
\end {figure}

\begin{figure}
\centering
\begin{picture} (150,200) (60,0)
\includegraphics[width=0.49\textwidth]{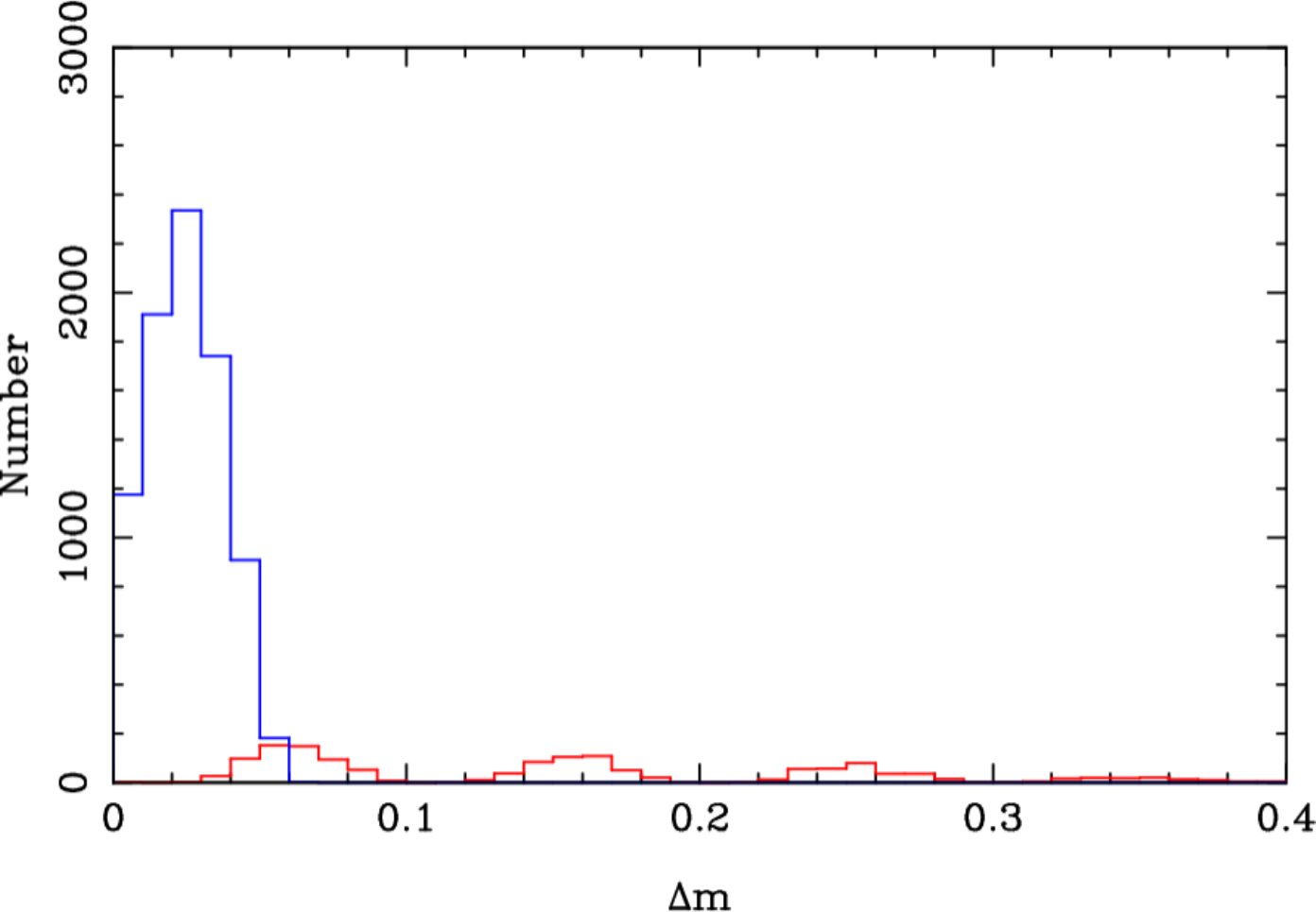}
\end{picture}
\caption{Histograms of changes in magnitude for a sample of light curves
 generated from the self organizing accretion disc model.  Blue and red
 histograms are for brightening and faintening changes in magnitude
 respectively.}
\label{fig16}
\end {figure}

The idea behind this paper is to test the prediction that if stellar
mass primordial black holes make up a significant fraction of the dark
matter, then the resulting optical depth to microlensing on a cosmological
scale will be sufficient to create a caustic web which when traversed by a
quasar accretion disc will result in characteristic caustic crossing
events in the quasar light curve.  For low redshift nearby sources any
microlensing events will be isolated and rare, and take the well-known
form of a bell-shaped Paczy\'{n}ski profile.  Such a feature in a quasar
light curve would be hard to identify unambiguously, and would not provide
an adequate test for the presence of lenses in the form of primordial
black holes.  However, in a $\Lambda$CDM universe caustics start to form
at around an optical depth $\tau \sim 0.1$ corresponding to a redshift
$z \sim 1.3$, leading to non-linear amplification of the quasar light and
characteristic structures in the quasar light curves.

Much of this paper has been devoted towards identifying features in quasar
light curves which are consistent with the trajectory of the quasar across
a web of caustics generated by a cosmological distribution of stellar mass
lenses.  In this context it is worth considering modes of intrinsic
variation which might mimic microlensing.  An important recent development
in the understanding of quasar structure has been reverberation mapping of
quasar accretion discs \citep{m18}.  This technique is based on a model of
quasar variabilty from \cite{c07} where central disturbances in the
nucleus are reprocessed and propagated through an accretion disc where the
temperature gradient is reflected in a matching colour gradient.  The
procedure adopted by \cite{m18} is to use the time lag between the arrival
of the central disturbance in different photometric passbands to measure
the size of the accretion disc.  The typical time lags are measured to be
a few days, roughly corresponding to light travel time between the colour
zones in the disc, with variations in the red lagging behind those in the
blue. This mode of variation would appear to be at odds with the analysis
of quasar light curves from Field 287 and Stripe 82 samples.  As can be
seen in Fig.~\ref{fig13}, time lags between photometric passbands are
typically of the order of a year, and are almost always characterised by
brightening in the red followed by more rapid brightening in the blue.
This seems to be a different mechanism to the intrinsic changes observed
in reverberation mapping projects.

The light curves containing candidate caustic crosssing events illustrated
in this paper should be seen as the identification of typical microlensing
structures rather than a statistically meaningful sample.  Although
ideally such a search procedure would be implemented with an objectively
defined algorithm, it became clear that the target light curves contained
too many ill-defined parameters for this to be feasible.    However, as
the idea was to investigate the possibility that such candidate caustic
crossing events were present in the data at all, a careful visual
inspection of each multicolour light curve turned out to give very good
results.  The two surveys which were used to identify the microlensing
events were significantly different in composition.  The main strength of
the Field 287 survey was the 26 year length of the light curves with
uniform uninterrupted yearly observations.  This fitted in well with the
characteristic timescale of 10 to 20 years for caustic crossing events
\citep{h07}.  By contrast, the Stripe 82 survey is made up of light curves
mostly covering 8 years, but in many cases with missing epochs.  Its two
great advantages are the 10,000 light curves it contains, in contrast to
1000 for the Field 287 sample, and the five photometric passbands of $u$,
$g$, $r$, $i$ and $z$ as opposed to just $B_J$ and $R$ for Field 287.

Although the search for caustic crossing events is not in any way
complete, there are aspects of microlensing which can be tested
statistically, and which can rule out modes of intrinsic changes in the
luminosity of the accretion disc.  The histogram in the top panel of
Fig.~\ref{fig12} is based on the statistics of increasing and decreasing
yearly magnitude increments for all 1033 quasars in the Field 287 sample.
The strong symmetry between increasing and decreasing changes in magnitude
puts tight constraints on the mechanism underlying the changes in quasar
brightness.  There have been a number of attempts to parametrise the
variations seen in quasar light curves, notably as a damped random walk
(DRW) process.  This approach has been implemented  by \cite{k10} who
propose an exponential covariance matrix of the form

\begin{equation}
%A = \frac{u^2 + 2}{u \sqrt{u^2+4}}
S_{ij} = \sigma^2 \exp (-|t_i-t_j|/\tau)
\label{eqn2}
\end{equation}

\noindent to model quasar light curves between epochs $t_i$ and $t_j$,
where $\sigma$ defines the amplitude and $\tau$ is the damping timescale.
This formulation was then used by \cite{m10} to model the time variability
of quasars in the SDSS Stripe 82.

This parametrisation of the variations in quasar light curves is
illustrated by the histogram in Fig.~\ref{fig14} using light curves
generated by Eq.~\ref{eqn2}, which has a similar structure to those in
the top panel of Fig.~\ref{fig12} and supports the idea that quasar light
curve variations may be consistent with a DRW process \citep{m10}.  This
parameterisation of the variations as statistically symmetric, combined
with the observational evidence in Fig.~\ref{fig12} puts strong
constraints on any physically motivated model of quasar variability, and
is inconsistent with the disc instability and starburst models discussed
in Section~\ref{res}.  Although there may be some room for asymmetry on
short timescales {$\lesssim 1$ year}, the expectation for a DRW process is
that the magnitude variations will be symmetric in time.  Given the
complex physical structure of an accretion disc it is perhaps not
surprising that models predicting symmetric variations in luminosity are
not prominent in the literature.  However, the more geometric nature of
microlensing variations avoid most of these problems and would appear to
provide a satisfactory model for the observed symmetry of variation,
although intrinsic modes of variation cannot be ruled out.

There is still no clear consensus on the nature of a physically
motivated model for intrinsic variations in quasar light on a timescale
of years.  The starburst model of \cite{t92} was taken seriously when
first published, but there has been little evidence to support it since
then.  The thermal reprocessing model of \cite{c07}, developed by
\cite{m18}, seems to work well for accretion disc variations on
timescales of days, but does not seem to be applicable on longer
timescales.  A promising mode of accretion disc variability proposed by
\cite{m94} and developed by \cite{t95} produces variations on the
accretion timescale from a self organizing accretion disc.  In this model
accretion causes matter to build up in the accretion disc, moving towards
the centre in a series of avalanches of various sizes.  Using the
description given by \cite{m94} to simulate the self organizing accretion
disc, a typical light curve generated by this process is shown in
Fig.~\ref{fig15}, and Fig.~\ref{fig16} shows histograms for brightening
and faintening changes in luminosity for a large sample of simulated light
curves.  It can be seen that the two histograms are quite different, with
the brightening changes largely confined to many small increases as
material slowly accretes onto the disc, whereas the disc becomes fainter
in much larger jumps as debris periodically cascades into the centre.

A further insight into quasar variability mechanisms can be obtained by
considering the symmetry of variations in accretion discs with a
significant colour gradient.  The bottom panel of Fig.~\ref{fig12} shows
histograms of yearly brightening magnitudes for blue and red passband
light curves, where the red histogram has been normalised to the blue
numbers to take account of the smaller number of complete red passband
light curves.  Similar histograms for faintening magnitudes show the same
structure.  It will be seen that relatively speaking the blue light
changes by larger increments than the red.  In a disc with a colour
gradient from blue in the centre changing to red in the outer parts this
would favour models where a central disturbance in the blue part of the
disc propagates outwards and decays towards the redder outer area.  In
disc instabilitiy models, a disturbance in the outer part of the disc,
caused perhaps by infalling debris, results in a red flare followed by a
confused response as the disturbance propagates through the remainder of
the disc. It would appear that models of this type are not favoured by the
histograms in the bottom panel of Fig.~\ref{fig12}.

The expectation for the statistics of microlensing variations is closer to
the central disturbance model rather than disc instabilty, as the central
blue compact core will be more strongly microlensed than the more diffuse
outer parts of the accretion disc.  There will however be a significant
difference in the timewise structure of the light curves.  Microlensing
will tend to produce an initial but relatively gentle rise in the red, as
the large diffuse area of red light starts to be microlensed.  This will
be followed by a much sharper rise in blue light as the compact blue
nucleus is microlensed.  This mode of variation is very distinctive, and
is indeed a feature of the simple caustic crossing simulated in
Figs~\ref{fig3} and ~\ref{fig4}.

The bottom panel of Fig.~\ref{fig12} illusrates another little discussed
statistical feature of colour changes in quasar light curves.  The rapid
rises and falls in blue light compared with the red light has a
geometrical explanation illustrated in Fig~\ref{fig4}, with observational
examples in Figs~\ref{fig6} and~\ref{fig7}.  It is hard to imagine a mode
of variation in accretion disc light which could mimic this, and certainly
no currently discussed accretion disc model has these properties.  This
includes the time lags in reverberation mapping measurements which are
typically on the order of a few days and are quite distinct from caustic
crossing events.

It is important to point out that these chromatic effects will only be
seen in quasar light curves when the visible part of the accretion disc
is larger than the Einstein radius of the lens.  Otherwise, any colour
gradient will not be resolved, and variations in light due to microlensing
will be achromatic.  This is consistent with observations of quasar light
curves which are largely characterised by achromatic variations.

\section{Conclusions}

The idea behind this paper has been to test the proposal that stellar mass
primordial black holes make up a significant fraction of the dark matter
by looking for evidence of caustic crossings in quasar light curves.  The
first part of the paper uses microlensing simulations to define the
expected features in quasar light curves produced by a cosmological
distribution of stellar mass primordial black holes making up the dark
matter.  These features are distinctive, and are not easily confused with
modes of variation in accretion disc luminosity.  The features in quasar
light curves which are associated with microlensing include the
characteristic cuspy structures resulting from caustic crossings and the
gradual rise in brightness in the red, followed by a sharper rise in the
blue as the colour gradient in an accretion disc is traversed by a
microlens. In addition, the statistics of yearly increase and decrease of
brightness in quasar light curves is symmetrical, as expected from the
geometrical nature of gravitational lensing.  The case is made that this
is not a property shared by current models of changes in accretion disc
luminosity.  These observations do not necessarily mean that all quasar
variability is caused by microlensing, but that if a substantial fraction
of dark matter is in the form of stellar mass primordial black, then the
expected microlensing features in quasar light curves are actually
observed.

\section*{Data Availability}

The data upon which this paper is based are all publicly available and are
referenced in the text, with footnotes to indicate online archives where
appropriate.

%%%%%%%%%%%%%%%%%%%%%%%%%%%%%%%%%%%%%%%%%%%%%%%%%%

% Don't change these lines
\bsp	% typesetting comment
\label{lastpage}
\end{document}